\documentclass[prb,superscriptaddress,showpacs,floatfix]{revtex4}
\usepackage{amssymb}
\usepackage{amsmath}
\usepackage{graphicx}

\begin{document}

\title{Pseudospin vortex-antivortex states with interwoven spin textures in
double layer quantum Hall systems }
\author{J. Bourassa}
\affiliation{D\'{e}partement de physique, Universit\'{e} de Sherbrooke, Sherbrooke, Qu%
\'{e}bec, Canada, J1K 2R1}
\author{B. Roostaei}
\affiliation{Department of Physics, University of Oklahoma, Norman, Oklahoma 730019-0390}
\author{R. C\^{o}t\'{e}}
\affiliation{D\'{e}partement de physique, Universit\'{e} de Sherbrooke, Sherbrooke, Qu%
\'{e}bec, Canada, J1K 2R1}
\email{Rene.Cote@Usherbrooke.ca}
\author{H. A. Fertig}
\affiliation{Department of Physics, Indiana University, Bloomington, Indiana 47405}
\author{K. Mullen}
\affiliation{Department of Physics, University of Oklahoma, Norman, Oklahoma 730019-0390}

\begin{abstract}
Recent experiments on strongly correlated bilayer quantum Hall systems
strongly suggest that, contrary to the usual assumption, the electron spin
degree of freedom is not completely frozen either in the quantum Hall or in
the compressibles states that occur at filling factor $\nu =1.$ These
experiments imply that the quasiparticles at $\nu =1$ could have both spin
and pseudospin textures i.e. they could be CP$^{3}$ skyrmions. Using a
microscopic unrestricted Hartree-Fock approximation, we compute the energy
of several crystal states with spin, pseudospin and mixed spin-pseudospin
textures around $\nu =1$ as a function of interlayer separation $d$ for
different values of tunneling ($\Delta _{SAS}$) , Zeeman ($\Delta _{Z}$),
and bias ($\Delta _{b}$) energies. We show that in some range of these
parameters, crystal states involving a certain amount of spin depolarization
have lower energy than the fully spin polarized crystals. We study this
depolarization dependence on $d,\Delta _{SAS},\Delta _{Z}$ and $\Delta _{b}$
and discuss how it can lead to the fast NMR relaxation rate observed
experimentally.
\end{abstract}

\date{\today}
\maketitle

\section{Introduction}

It is now well established that the two-dimensional electron gas (2DEG) in a
double-quantum-well system (DQWS) at filling factor $\nu =1$ has a broken
symmetry ground state that can be described as either an easy-plane
pseudospin ferromagnet or as an excitonic superfluid. The ferromagnetic
state (in the pseudospin language) has finite interlayer coherence even in
the absence of tunneling if the interlayer separation $d$ is lower than a
critical layer separation $d_{c}\approx 1.2\ell $ where $\ell =\sqrt{\hslash
c/eB}$ is the magnetic length. This is an incompressible state supporting a
quantum Hall effect (QHE). The phase diagram and physical properties of this
state have been extensively studied over the past fifteen years (for a
review, see Refs. \onlinecite{dassarmalivre},\onlinecite{ezawalivre}).

In most studies of the bilayer coherent states at or near filling factor $%
\nu =1$, it is generally assumed that, due to the strong magnetic field, the
ground state is fully spin polarized. The spin degrees of freedom could thus
be left out of the analysis. Recent experiments, however, cast some doubt on
the validity of this assumption. These experiments include the measurement%
\cite{sawadacp3} of the dependence of the activation energy of the bilayer
quantum Hall state at $\nu =1$ on an in-plane field, and the measurement\cite%
{spielmanprl, kumadaprl} of the nuclear magnetic relaxation (NMR) time $%
T_{1} $ near $\nu =1$.

The monolayer quantum Hall state at $\nu =1$ is a spin ferromagnet. In the
absence of Zeeman coupling, the lowest-energy charged excitation is a
spin-textured topological object called a Skyrmion\cite{sondhi,
barret,schmeller}. Measurement of the activation energy shows an \textit{%
increase} of the energy gap when the magnetic field is tilted away from the $%
z$ axis at $\nu =1$. This is easily understood since keeping the filling
factor constant means increasing the magnetic field, so the increase in
activation energy reflects the increase in the Zeeman energy cost of the
skyrmions. By contrast, measurement\cite{murphy} of the activation energy in
the bilayer quantum Hall state at $\nu =1$ shows a strong \textit{decrease}
of this energy with tilting until some critical angle $\theta _{c}$ above
which the activation energy ceases to depend on $\theta .$ This anomalous
behaviour of the activation energy was interpreted as a change in the ground
state of the system at $\theta _{c}$ due to a the occurence of a
commensurate-incommensurate transition.\cite{kunyangci} The lowest-energy
charged excitation of the spin-polarized bilayer system at $\nu =1$ is a 
\textit{bimeron}, i.e. a skyrmion in the pseudospin field. A Hartree-Fock
calculation of the behavior of the bimeron energy in tilted magnetic field
reproduces qualitatively the features found in the experiment.\cite%
{breycristalbimerons}

Sawada \textit{et al.}\cite{sawadacp3} measured the activation energy of the
2DEG in a DQWS as a function of a parallel magnetic field and electrical
bias between the layers. They define the imbalance parameter $\sigma
=(n_{L}-n_{R})/(n_{L}+n_{R})$ where $n_{R(L)}$ is the density in the
right(left) layer of the DQWS. At $\sigma =0$, the activation energy showed
the behavior expected for pseudospin-skyrmions (i.e., bimerons) while at $%
\sigma =1$, where all electrons reside in one well, the behavior was that
expected for a spin-skyrmion. Sawada \textit{et al.} found a continuous
evolution from pseudospin-skyrmion to spin-skyrmion as the imbalance
parameter was increased in various samples with tunneling energies ranging
from $\Delta _{SAS}=1$K to $\Delta _{SAS}=33$K. They concluded that the
excited quasiparticles must contain both spin and pseudospin flips in order
to explain their results. In particular, the behavior of the activation
energy could not be explained by a level crossing between skyrmion and a
bimeron excitations. Instead, the bimeron excitation, at the balance point, 
\textit{continuously} transformed into a spin-skyrmion at high bias. This
suggests the quasiparticles at these biases may be some object that
interpolates between the two types of skyrmions. Such objects have been
studied in the field theoretic literature by Ghosh and Rajaraman\cite%
{rajaramancp3}, and more recently by Ezawa and Tsitsishvili\cite{ezawasu4},
who dubbed these objects CP$^{3}$ skyrmions. In this last work, good
agreement between theoretical calculations and the measurements of Sawada 
\textit{et al. } were obtained.

Another set of experiments confirming the necessity to take into account
spin depolarization in the ground state of the 2DEG around $\nu =1$ are
those of Spielman \textit{et al.}\cite{spielmanprl} and by Kumada \textit{et
al.}\cite{kumadaprl}. In these experiments, the NMR$\ $relaxation time $%
T_{1} $ is measured as a function of filling factor. (In Ref. %
\onlinecite{kumadaprl}, this is also done as a function of an electrical
bias.) The behavior of $T_{1}$ seen in these experiments is reminescent of
that measured\cite{tyckonmr} in monolayer quantum Hall systems where the
relaxation rate $T_{1}^{-1}$ increases when $\nu $ deviates from $1$. A
possible explanation \cite{cotegirvinprl} involves the inclusion of
skyrmions in the groundstate when the system is doped away from $\nu =1$. A
single skyrmion has it spin aligned with the Zeeman field at infinity,
reversed at at the center of the skyrmions, and has nonzero $XY$ spin
components at intermediate distances in a vortex-like configuration. For $%
|\nu -1|>0$, the finite density of these objects is expected to condense
into a crystal \cite{breyprl}. The quantum mean-field energy of this crystal
is independent of the angle $\varphi $ which defines the global orientation
of the $XY$ spin components. C\^{o}t\'{e} \textit{et al.}\cite{cotegirvinprl}
showed that this extra $U(1)$ degree of freedom leads to broken symmetry and
hence to a spin wave mode that remains gapless in the presence of the Zeeman
field. It is the existence of this extra gapless spin mode in the crystal
phase (and possibly in some overdamped form in a Skyrme liquid state) that
is believed to be responsible for the rapid nuclear spin relaxation observed
in the experiments.

Our goal in this paper is to show that crystal states with some amount of
spin depolarization due either to spin-skyrmions or CP$^{3}$ skyrmions exist
around $\nu =1$, with lower energy than crystal states with maximal spin
polarization. We show this by comparing the energy of several crystal states
in the Hartree-Fock approximation. We study the spin and pseudospin textures
of these states as the interlayer separation, the Zeeman, tunnel coupling or
the electrical bias are varied. Because CP$^{3}$ crystals must have a
gapless spin wave mode, just as the Skyrme crystal in a monolayer QH system
has, this crystal state could be responsible for the fast NMR relaxation
rate seen in the experiments of Spielman \textit{et al.}\cite{spielmanprl}
and Kumada \textit{et al.}\cite{kumadaprl}. In addition, these Hartree-Fock
calculations provide a check on the field theoretic calculations that have
been used on the CP$^{3}$ state\cite{rajaramancp3,ezawasu4}. In the latter
it is assumed that the many body wavefunction can be represented by a local
4-spinor with constant magnitude. While Hartree-Fock introduces its own
approximations by using a different approach we may hope to better
understand the true many body state.

Our paper is organized as follow. The model hamiltonian and the Hartree-Fock
formalism needed to compute the CP$^{3}$\ Skyrme crystal are introduced in
Sec. II. In Sec. III, we introduce the CP$^{3}$ skyrmion and discuss its
limiting forms: spin-skyrmion and pseudospin-skyrmion. Our numerical results
are presented in Sec. IV. We discuss the relevance of our results to the
experiments of Spielman \textit{et al.}\cite{spielmanprl} and Kumada \textit{%
et al.}\cite{kumadaprl} in Sec. V.

\section{Model hamiltonian and Hartree-Fock formalism}

\subsection{Hamiltonian of the 2DEG}

We consider a symmetric double-quantum-well system in a magnetic field $%
\mathbf{B}=B\widehat{\mathbf{z}}$ and submitted to an electrical bias $%
\Delta _{b}=E_{R}-E_{L}$ where $E_{R\left( L\right) }$ are the subband
energies in each layer (right and left) in the absence of magnetic field and
tunneling. For the sake of limiting the number of parameters characterizing
our DQWS, we make a narrow well approximation, i.e. we assume that the width 
$b$ of the wells is small ($b<<\ell $) and treat interlayer hopping in a
tight-binding approximation. The single-particle problem is then
characterized by the separation $d$ (from center to center) between the
wells and the tunneling integral $t=\Delta _{SAS}/2$. In the Landau gauge
where the potential vector is taken as $\mathbf{A}=\left( 0,Bx,0\right) $,
the Hamiltonian $H_{0}$ of the non-interacting 2DEG is given by%
\begin{eqnarray}
H_{0} &=&\sum_{X,j,\alpha }E_{j,\alpha }c_{X,j,\alpha }^{\dagger
}c_{X,j,\alpha }  \label{ho} \\
&&-t\sum_{X,\alpha }\left( c_{X,R,\alpha }^{\dagger }c_{X,L,\alpha
}+c_{X,L,\alpha }^{\dagger }c_{X,R,\alpha }\right) .  \notag
\end{eqnarray}%
In Eq. (\ref{ho}), $c_{X,j,\alpha }^{\dagger }$ is an operator that creates
an electron with guiding center $X$ in well $j=R,L$ with spin index $\alpha
=\pm 1.$ We work in the strong quantum limit where we assume that Landau
level mixing is negligible and only the Landau level $N=0$ needs to be
considered. The energies $E_{j,\alpha }$ are defined by $E_{R,a}=\Delta
_{b}/2+\alpha \Delta _{Z}/2$ and $E_{L,a}=-\Delta _{b}/2+\alpha \Delta
_{Z}/2 $ where $\Delta _{Z}=g^{\ast }\mu _{B}B$ is the Zeeman energy, $%
g^{\ast }$ is the effective gyromagnetic factor and $\mu _{B}$ is the Bohr
magneton.

We describe the various phases of the electrons in the DQWS by the set of
average values $\left\{ \left\langle \rho _{i,j}^{\alpha ,\beta }(\mathbf{q}%
)\right\rangle \right\} $ where $\rho _{i,j}^{\alpha ,\beta }(\mathbf{q})$
is an operator that we define\cite{cotemethode} by 
\begin{equation*}
\rho _{i,j}^{\alpha ,\beta }(\mathbf{q})=\frac{1}{N_{\varphi }}%
\sum_{X}e^{-iq_{x}X+iq_{x}q_{y}\ell ^{2}/2}\ c_{i,\alpha ,X}^{\dagger
}c_{j,\beta ,X-q_{y}\ell ^{2}},
\end{equation*}%
where the Landau level degeneracy is $N_{\varphi }=S/2\pi \ell ^{2}$ (with $%
S $ the area of the 2DEG). We explain the physical meaning of these
operators below.

In the Hartree-Fock approximation, the Hamiltonian of the interacting 2DEG
in the DQWS is given by

\begin{eqnarray*}
H_{HF}&=& N_{\phi }\sum_{i,\alpha }\widetilde{E}_{\alpha ,i}\rho
_{i,i}^{\alpha ,\alpha }\left( 0\right)  \label{hhf} \\
&&-N_{\phi }t\sum_{\alpha }\left[ \rho _{R,L}^{\alpha ,\alpha }\left(
0\right) +\rho _{L,R}^{\alpha ,\alpha }\left( 0\right) \right] \\
&&+N_{\phi }\left( \frac{e^{2}}{\kappa \ell }\right) \sum_{\alpha ,\beta
}\sum_{i,j}\sum_{\mathbf{q\neq }0}H_{i,j}\left( q\right) \left\langle \rho
_{i,i}^{\alpha ,\alpha }(-\mathbf{q})\right\rangle \rho _{j,j}^{\beta ,\beta
}(\mathbf{q}) \\
&& -N_{\phi }\left( \frac{e^{2}}{\kappa \ell }\right) \sum_{\alpha ,\beta
}\sum_{i,j}\sum_{\mathbf{q}}X_{i,j}\left( q\right) \left\langle \rho
_{i,j}^{\alpha ,\beta }(-\mathbf{q})\right\rangle \rho _{j,i}^{\beta ,\alpha
}(\mathbf{q}),
\end{eqnarray*}

where the renormalized single-particle energies $\widetilde{E}_{\alpha ,i}$
are defined by

\begin{equation}
\widetilde{E}_{\alpha ,i}=E_{i,\alpha }+\left( \frac{e^{2}}{\kappa \ell }%
\right) \left[ \frac{\nu }{2}\left( \frac{d}{\ell }\right) -\frac{d\nu _{%
\overline{i}}}{\ell }\right] .  \label{enia}
\end{equation}%
In Eq. (\ref{enia}), $\nu _{\overline{R}}=\sum_{\alpha }\nu _{L,\alpha }$
and $\nu _{\overline{L}}=\sum_{\alpha }\nu _{R,\alpha }$ where $\nu
_{i,\alpha }$ is the filling factor for state $\left( i,\alpha \right) $ and 
$\nu =\sum_{j,\alpha }\nu _{j,\alpha }$ is the total filling factor of the
2DEG. The Hartree and Fock intrawell and interwell interactions are defined
by%
\begin{eqnarray*}
H_{i,i}\left( q\right) &=&H\left( q\right) =\frac{1}{q\ell }e^{-q^{2}\ell
^{2}}, \\
H_{i\neq j}\left( q\right) &=&\widetilde{H}\left( q\right) =\frac{1}{q\ell }%
e^{-q^{2}\ell ^{2}}e^{-qd}, \\
X_{i,i}\left( q\right) &=&X\left( q\right) =\int_{0}^{+\infty
}dye^{-y^{2}/2}J_{0}\left( q\ell y\right) , \\
X_{i\neq j}\left( q\right) &=&\widetilde{X}\left( q\right)
=\int_{0}^{+\infty }dye^{-y^{2}/2}e^{-dy/\ell }J_{0}\left( q\ell y\right) .
\end{eqnarray*}

\subsection{Calculation of the order parameters $\left\{ \left\langle 
\protect\rho _{i,j}^{\protect\alpha ,\protect\beta }(\mathbf{q}%
)\right\rangle \right\} $ of the crystal phases\qquad}

To simplify our notation, we now define the four states $1,2,3,4\equiv
\left( R,+\right) ,(R,-),(L,+),(L,-)$ and write the order parameters $%
\left\langle \rho _{i,j}^{\alpha ,\beta }(\mathbf{q})\right\rangle $ as $%
\left\langle \rho _{i,j}(\mathbf{q})\right\rangle $. From now on, the
indices $i,j$ will run from $1$ to $4$. The average values $\left\langle
\rho _{i,j}(\mathbf{q})\right\rangle $ are obtained by computing the
single-particle Green's function 
\begin{equation*}
G_{i,j}\left( X,X^{\prime },\tau \right) =-\left\langle Tc_{i,X}\left( \tau
\right) c_{j,X^{\prime }}^{\dagger }\left( 0\right) \right\rangle ,
\end{equation*}%
whose Fourier transform we define as 
\begin{equation*}
G_{i,j}\left( \mathbf{q,}\tau \right) =\frac{1}{N_{\phi }}\sum_{X,X^{\prime
}}e^{-\frac{i}{2}q_{x}\left( X+X^{\prime }\right) }\delta _{X,X^{\prime
}-q_{y}l^{2}}G_{i,j}\left( X,X^{\prime },\tau \right) ,
\end{equation*}%
so that $G_{i,j}\left( \mathbf{q,}\tau =0^{-}\right) =\left\langle \rho
_{j,i}\left( \mathbf{q}\right) \right\rangle $. In a homogeneous phase, only 
$\left\{ \left\langle \rho _{i,j}^{\alpha ,\beta }(\mathbf{q=0}%
)\right\rangle \right\} $ are nonzero while, in a crystal, $\left\langle
\rho _{j,i}\left( \mathbf{q}\right) \right\rangle \neq 0$ only if $\mathbf{%
q\in }\left\{ \mathbf{G}\right\} $ where $\left\{ \mathbf{G}\right\} $ is
the set of reciprocal lattice vectors of the crystal.

In our numerical calculation, we consider a finite number $N$ of reciprocal
latttice vectors $\left( \mathbf{G}_{1},\mathbf{G}_{2},...,\mathbf{G}%
_{N}\right) $. Defining the column vectors%
\begin{equation*}
\overline{G}_{i,j}=\left( 
\begin{array}{c}
G_{i,j}\left( \mathbf{G}_{1},\omega _{n}\right) \\ 
G_{i,j}\left( \mathbf{G}_{2},\omega _{n}\right) \\ 
\vdots \\ 
G_{i,j}\left( \mathbf{G}_{N},\omega _{n}\right)%
\end{array}%
\right) ,
\end{equation*}%
where $\omega _{n}$ is a Matsubara frequency, and the vectors 
\begin{equation*}
\overline{B}=\left( 
\begin{array}{c}
1 \\ 
0 \\ 
\vdots \\ 
0%
\end{array}%
\right) ,\overline{0}=\left( 
\begin{array}{c}
0 \\ 
0 \\ 
\vdots \\ 
0%
\end{array}%
\right) ,
\end{equation*}%
along with the $4N\times 4$ matrices%
\begin{equation*}
\overline{\overline{G}}=\left( 
\begin{array}{cccc}
\overline{G}_{1,1} & \overline{G}_{1,2} & \overline{G}_{1,3} & \overline{G}%
_{1,4} \\ 
\overline{G}_{2,1} & \overline{G}_{2,2} & \overline{G}_{2,3} & \overline{G}%
_{2,4} \\ 
\overline{G}_{3,1} & \overline{G}_{3,2} & \overline{G}_{3,3} & \overline{G}%
_{3,4} \\ 
\overline{G}_{4,1} & \overline{G}_{4,2} & \overline{G}_{4,3} & \overline{G}%
_{4,4}%
\end{array}%
\right) ,
\end{equation*}%
and%
\begin{equation*}
\overline{\overline{C}}=\hslash \left( 
\begin{array}{cccc}
\overline{B} & \overline{0} & \overline{0} & \overline{0} \\ 
\overline{0} & \overline{B} & \overline{0} & \overline{0} \\ 
\overline{0} & \overline{0} & \overline{B} & \overline{0} \\ 
\overline{0} & \overline{0} & \overline{0} & \overline{B}%
\end{array}%
\right) ,
\end{equation*}%
we find that the Hartree-Fock equation of motion for the single-particle
Green's function matrix $\overline{\overline{G}}$ can be written in a matrix
form as%
\begin{equation}
\left( i\hslash \omega _{n}+\mu \right) \overline{\overline{G}}-\overline{%
\overline{A}}~\overline{\overline{G}}=\overline{\overline{C}},
\label{motion}
\end{equation}%
where $\overline{\overline{A}}$ is the $4N\times 4N$ hermitian matrix

\begin{widetext}
\begin{equation*}
\overline{\overline{A}}=\left( 
\begin{array}{cccc}
\Lambda _{1}\gamma  & -X\left\langle \rho _{2,1}\right\rangle \gamma  & 
-t\delta _{\mathbf{G,G}^{\prime }}-\widetilde{X}\left\langle \rho
_{3,1}\right\rangle \gamma  & -\widetilde{X}\left\langle \rho
_{4,1}\right\rangle \gamma  \\ 
-X\left\langle \rho _{1,2}\right\rangle \gamma  & \Lambda _{2} \gamma & -%
\widetilde{X}\left\langle \rho _{3,2}\right\rangle \gamma  & -t\delta _{%
\mathbf{G,G}^{\prime }}-\widetilde{X}\left\langle \rho _{4,2}\right\rangle
\gamma  \\ 
-t\delta _{\mathbf{G,G}^{\prime }}-\widetilde{X}\left\langle \rho
_{1,3}\right\rangle \gamma  & -\widetilde{X}\left\langle \rho
_{2,3}\right\rangle \gamma  & \Lambda _{3}\gamma  & -X\left\langle \rho
_{4,3}\right\rangle \gamma  \\ 
-\widetilde{X}\left\langle \rho _{1,4}\right\rangle \gamma  & -t\delta _{%
\mathbf{G,G}^{\prime }}-\widetilde{X}\left\langle \rho _{2,4}\right\rangle
\gamma  & -X\left\langle \rho _{3,4}\right\rangle \gamma  & \Lambda
_{4}\gamma 
\end{array}%
\right) 
\end{equation*}%
\end{widetext}
with $\gamma =e^{-i\left( \mathbf{G}\times \mathbf{G}^{\prime }\right) \cdot 
\widehat{\mathbf{z}}\ell ^{2}/2},$%
\begin{equation*}
\Lambda _{i}=\widetilde{E}_{i}\delta _{\mathbf{G,G}^{\prime }}+\Upsilon
_{i}-X\left\langle \rho _{i,i}\right\rangle ,
\end{equation*}%
and 
\smallskip
\smallskip
\begin{equation}
\Upsilon _{i}=\Upsilon _{i}\left( \mathbf{G-G}^{\prime }\right)
=\sum_{j}H_{j,i}\left( \mathbf{G-G}^{\prime }\right) \left\langle \rho
_{j,j}\left( \mathbf{G-G}^{\prime }\right) \right\rangle .
\end{equation}%
In the definition of $\overline{\overline{A}}$, the symbols $\Lambda _{i},X$
and $\widetilde{X}$ stand for $\Lambda _{i}\left( \mathbf{G-G}^{\prime
}\right) ,X\left( \mathbf{G-G}^{\prime }\right) ,\widetilde{X}\left( \mathbf{%
G-G}^{\prime }\right) $ and the quantity $\left\langle \rho
_{i,j}\right\rangle $ for $\left\langle \rho _{i,j}\left( \mathbf{G-G}%
^{\prime }\right) \right\rangle .$

The $\left\langle \rho _{i,j}\left( \mathbf{G}\right) \right\rangle ^{\prime
}s$ are found by solving numerically the self-consistent equation of motion
given by Eq. (\ref{motion}). This equation has many solutions representing
the local minima of the Hartree-Fock energy given by%
\begin{eqnarray*}
\frac{E_{HF}}{N} &=&\frac{1}{\nu }\sum_{i}E_{i}\left\langle \rho
_{i,i}\left( 0\right) \right\rangle +\left( \frac{e^{2}}{\kappa \ell }%
\right) \frac{d}{\ell }\frac{\left( \nu _{R}-\nu _{L}\right) ^{2}}{4\nu } \\
&&-\frac{1}{\nu }\frac{\Delta _{SAS}}{2}\sum_{i=1,2}\left[ \left\langle \rho
_{i,i+2}\left( 0\right) \right\rangle +\left\langle \rho _{i+2,i}\left(
0\right) \right\rangle \right] \\
&&+\frac{1}{2\nu }\left( \frac{e^{2}}{\kappa \ell }\right) \sum_{i,j}\sum_{%
\mathbf{G\neq 0}}H_{i,j}\left( \mathbf{G}\right) \left\langle \rho
_{i,i}\left( -\mathbf{G}\right) \right\rangle \left\langle \rho _{j,j}\left( 
\mathbf{G}\right) \right\rangle \\
&&-\frac{1}{2\nu }\left( \frac{e^{2}}{\kappa \ell }\right) \sum_{i,j}\sum_{%
\mathbf{G}}X_{i,j}\left( \mathbf{G}\right) \left\vert \left\langle \rho
_{i,j}\left( \mathbf{G}\right) \right\rangle \right\vert ^{2}.
\end{eqnarray*}

There is no guarantee that a solution is an absolute minimum of $E_{HF}$.
Instead, we compare a finite number of likely solutions and choose the one
that minimizes the energy. The numerical scheme to solve Eq.(\ref{motion})
is described in more detail in Ref. \onlinecite{cotemethode}.

By definition, 
\begin{equation*}
\sum_{i}\left\langle \rho _{i,i}\left( 0\right) \right\rangle =\nu .
\end{equation*}%
This equation fixes the chemical potential $\mu $ in Eq.(\ref{motion}).

\subsection{Spin and pseudospin fields in the crystal phases}

In the Landau gauge and with the Hilbert space restricted to the first
Landau level only, an electronic state in a single quantum well system
(SQWS) is specified by the two-component spinor $c_{X}$ where 
\begin{equation*}
c_{X}=\left( 
\begin{array}{c}
c_{X,+} \\ 
c_{X,-}%
\end{array}%
\right) .
\end{equation*}%
Similarly, an electronic state in a \textit{spin-polarized} DQWS can be
described by mapping this two-level system into a spin $1/2$ system by using
the pseudo-spinor $c_{X}$ where%
\begin{equation*}
c_{X}=\left( 
\begin{array}{c}
c_{X,R} \\ 
c_{X,L}%
\end{array}%
\right) .
\end{equation*}%
For a four-level system (with states $j=1,2,3,4=\left( R,+\right)
,(R,-),(L,+),(L,-)$ as given above), an electronic state is specified by the
four-component spinor%
\begin{equation*}
c_{X}=\left( 
\begin{array}{c}
c_{X,1} \\ 
c_{X,2} \\ 
c_{X,3} \\ 
c_{X,4}%
\end{array}%
\right) .
\end{equation*}

The operators $\rho _{i,j}(\mathbf{q})$ that we introduced previously can be
mapped into the density operator, $\rho \left( \mathbf{q}\right) $, the spin
and pseudospin densities operators $S_{a}\left( \mathbf{q}\right) $ and $%
P_{a}\left( \mathbf{q}\right) $ (with $a=x,y,z$) and the 9 operators $%
R_{a,b}\left( \mathbf{q}\right) $ using the SU(4) algebra (as defined in
Ref. \onlinecite{ezawasu4})%
\begin{equation}
\rho (\mathbf{q})=\frac{1}{N_{\varphi }}\sum_{X}e^{-iq_{x}X+iq_{x}q_{y}\ell
^{2}/2}\ c_{X}^{\dagger }c_{X-q_{y}\ell ^{2}},  \label{a1}
\end{equation}%
\begin{equation}
S_{a}(\mathbf{q})=\frac{1}{N_{\varphi }}\sum_{X}e^{-iq_{x}X+iq_{x}q_{y}\ell
^{2}/2}\ c_{X}^{\dagger }\tau _{a}^{spin}c_{X-q_{y}\ell ^{2}},  \label{a2}
\end{equation}%
\begin{equation}
P_{a}(\mathbf{q})=\frac{1}{N_{\varphi }}\sum_{X}e^{-iq_{x}X+iq_{x}q_{y}\ell
^{2}/2}\ c_{X}^{\dagger }\tau _{a}^{ppin}c_{X-q_{y}\ell ^{2}},  \label{a3}
\end{equation}%
\begin{equation}
R_{a,b}(\mathbf{q})=\frac{1}{N_{\varphi }}\sum_{X}e^{-iq_{x}X+iq_{x}q_{y}%
\ell ^{2}/2}\ c_{X}^{\dagger }\tau _{a}^{spin}\tau _{b}^{ppin}c_{X-q_{y}\ell
^{2}},  \label{a4}
\end{equation}%
where the $4\times 4$ matrices $\tau _{a}^{spin}$ and $\tau _{a}^{ppin}$ are
defined by

\begin{equation*}
\tau _{a}^{spin}=\left( 
\begin{array}{cc}
\sigma _{a} & 0 \\ 
0 & \sigma _{a}%
\end{array}%
\right) ,
\end{equation*}%
with $\sigma _{a}$ a Pauli matrix, and by%
\begin{equation*}
\tau _{x}^{ppin}=\left( 
\begin{array}{cc}
0 & I \\ 
I & 0%
\end{array}%
\right) ,\tau _{y}^{ppin}=\left( 
\begin{array}{cc}
0 & -iI \\ 
iI & 0%
\end{array}%
\right) ,\tau _{z}^{ppin}=\left( 
\begin{array}{cc}
I & 0 \\ 
0 & -I%
\end{array}%
\right) ,
\end{equation*}%
where $I$ is the $2\times 2$ unit matrix.

From Eqs. (\ref{a1})-(\ref{a4}), it is easy to show that the electronic
densities in the right and left wells are given by

\begin{eqnarray*}
\left\langle \rho _{R}\left( \mathbf{q}\right) \right\rangle &=&\left\langle
\rho _{1,1}\left( \mathbf{q}\right) \right\rangle +\left\langle \rho
_{2,2}\left( \mathbf{q}\right) \right\rangle , \\
\left\langle \rho _{L}\left( \mathbf{q}\right) \right\rangle &=&\left\langle
\rho _{3,3}\left( \mathbf{q}\right) \right\rangle +\left\langle \rho
_{4,4}\left( \mathbf{q}\right) \right\rangle ,
\end{eqnarray*}%
with the total electronic density given by $\left\langle \rho \left( \mathbf{%
q}\right) \right\rangle =\left\langle \rho _{R}\left( \mathbf{q}\right)
\right\rangle +\left\langle \rho _{L}\left( \mathbf{q}\right) \right\rangle
. $

The spin densities in the right and left wells are given by%
\begin{eqnarray*}
\left\langle S_{x,R}\left( \mathbf{q}\right) \right\rangle &=&\Re
\left\langle \rho _{1,2}\left( \mathbf{q}\right) \right\rangle , \\
\left\langle S_{y,R}\left( \mathbf{q}\right) \right\rangle &=&\Im
\left\langle \rho _{1,2}\left( \mathbf{q}\right) \right\rangle ,
\end{eqnarray*}%
\begin{eqnarray*}
\left\langle S_{x,L}\left( \mathbf{q}\right) \right\rangle &=&\Re
\left\langle \rho _{3,4}\left( \mathbf{q}\right) \right\rangle , \\
\left\langle S_{y,L}\left( \mathbf{q}\right) \right\rangle &=&\Im
\left\langle \rho _{3,4}\left( \mathbf{q}\right) \right\rangle ,
\end{eqnarray*}%
and by%
\begin{eqnarray*}
\left\langle S_{z,R}\left( \mathbf{q}\right) \right\rangle &=&\frac{1}{2}%
\left[ \left\langle \rho _{1,1}\left( \mathbf{q}\right) \right\rangle
-\left\langle \rho _{2,2}\left( \mathbf{q}\right) \right\rangle \right] , \\
\left\langle S_{z,L}\left( \mathbf{q}\right) \right\rangle &=&\frac{1}{2}%
\left[ \left\langle \rho _{3,3}\left( \mathbf{q}\right) \right\rangle
-\left\langle \rho _{4,4}\left( \mathbf{q}\right) \right\rangle \right] .
\end{eqnarray*}

Finally, the pseudospin densities for the up (+) and down (-) spin
components are given by%
\begin{eqnarray*}
\left\langle P_{x,+}\left( \mathbf{q}\right) \right\rangle &=&\Re
\left\langle \rho _{1,3}\left( \mathbf{q}\right) \right\rangle , \\
\left\langle P_{y,+}\left( \mathbf{q}\right) \right\rangle &=&\Im
\left\langle \rho _{1,3}\left( \mathbf{q}\right) \right\rangle ,
\end{eqnarray*}%
\begin{eqnarray*}
\left\langle P_{x,-}\left( \mathbf{q}\right) \right\rangle &=&\Re
\left\langle \rho _{2,4}\left( \mathbf{q}\right) \right\rangle , \\
\left\langle P_{y,-}\left( \mathbf{q}\right) \right\rangle &=&\Im
\left\langle \rho _{2,4}\left( \mathbf{q}\right) \right\rangle ,
\end{eqnarray*}%
and by%
\begin{eqnarray*}
\left\langle P_{z,+}\left( \mathbf{q}\right) \right\rangle &=&\frac{1}{2}%
\left[ \left\langle \rho _{1,1}\left( \mathbf{q}\right) \right\rangle
-\left\langle \rho _{3,3}\left( \mathbf{q}\right) \right\rangle \right] , \\
\left\langle P_{z,-}\left( \mathbf{q}\right) \right\rangle &=&\frac{1}{2}%
\left[ \left\langle \rho _{2,2}\left( \mathbf{q}\right) \right\rangle
-\left\langle \rho _{4,4}\left( \mathbf{q}\right) \right\rangle \right] ,
\end{eqnarray*}%
with $P_{z}\left( \mathbf{q}\right) =P_{z,+}\left( \mathbf{q}\right)
+P_{z,-}\left( \mathbf{q}\right) $ the total pseudospin density.

Note that by definition $\left\langle \rho _{i,j}\left( \mathbf{q}\right)
\right\rangle =\left\langle \rho _{j,i}\left( -\mathbf{q}\right)
\right\rangle ^{\ast }.$ Also, $\left\langle \mathbf{S}\left( -\mathbf{q}%
\right) \right\rangle =\left\langle \mathbf{S}\left( \mathbf{q}\right)
\right\rangle ^{\ast }$ and $\left\langle \mathbf{P}\left( -\mathbf{q}%
\right) \right\rangle =\left\langle \mathbf{P}\left( \mathbf{q}\right)
\right\rangle ^{\ast }$. The four order parameters that are not related to
the electron, spin, or pseudospin densities are $\left\langle \rho
_{1,4}\left( \mathbf{q}\right) \right\rangle ,\left\langle \rho _{2,3}\left( 
\mathbf{q}\right) \right\rangle \mathbf{\ }$and their complex conjugates.
These densities involve average values of operators that flip both the spin
and the pseudospin.

The Hartree-Fock energy per electron can now be written as%
\begin{gather}
\frac{E_{HF}}{N}=\frac{\Delta _{b}}{\nu }\left\langle P_{z}\left( \mathbf{0}%
\right) \right\rangle -\frac{\Delta _{Z}}{\nu }\left\langle S_{z}\left( 
\mathbf{0}\right) \right\rangle -\frac{\Delta _{SAS}}{\nu }\left\langle
P_{x}\left( \mathbf{0}\right) \right\rangle  \notag \\
+\frac{1}{4\nu }\sum_{\mathbf{G}}\Upsilon _{1}\left( \mathbf{G}\right)
\left\vert \left\langle \rho \left( \mathbf{G}\right) \right\rangle
\right\vert ^{2}+\frac{1}{\nu }\sum_{\mathbf{G}}J_{z,1}\left( \mathbf{G}%
\right) \left\vert \left\langle P_{z}\left( \mathbf{G}\right) \right\rangle
\right\vert ^{2}  \notag \\
-\frac{1}{\nu }\sum_{\mathbf{G}}\sum_{a=R,L}X\left( \mathbf{G}\right)
\left\vert \left\langle \mathbf{S}_{a}\left( \mathbf{G}\right) \right\rangle
\right\vert ^{2}  \label{enerpseudos} \\
-\frac{1}{\nu }\sum_{\mathbf{G}}\sum_{\alpha =+,-}\widetilde{X}\left( 
\mathbf{G}\right) \left[ \left\vert \left\langle P_{x,\alpha }\left( \mathbf{%
G}\right) \right\rangle \right\vert ^{2}+\left\vert \left\langle P_{y,\alpha
}\left( \mathbf{G}\right) \right\rangle \right\vert ^{2}\right]  \notag \\
-\frac{1}{\nu }\sum_{\mathbf{G}}\widetilde{X}\left( \mathbf{G}\right) \left[
\left\vert \left\langle \rho _{1,4}\left( \mathbf{G}\right) \right\rangle
\right\vert ^{2}+\left\vert \left\langle \rho _{2,3}\left( \mathbf{G}\right)
\right\rangle \right\vert ^{2}\right] .  \notag
\end{gather}%
In Eq. (\ref{enerpseudos}), we have defined the interactions 
\begin{eqnarray*}
\Upsilon _{1}\left( \mathbf{G}\right) &=&H\left( \mathbf{G}\right) +%
\widetilde{H}\left( \mathbf{G}\right) -\frac{1}{2}X\left( \mathbf{G}\right) ,
\\
J_{z,1}\left( \mathbf{G}\right) &=&H\left( \mathbf{G}\right) -\widetilde{H}%
\left( \mathbf{G}\right) -\frac{1}{2}X\left( \mathbf{G}\right) .
\end{eqnarray*}

Because of the neutrality of the total system comprising the electrons and
the positive donors, we have%
\begin{eqnarray*}
\Upsilon _{1}\left( 0\right) &=&-\frac{1}{2}X\left( 0\right) , \\
J_{z,1}\left( 0\right) &=&\frac{d}{\ell }-\frac{1}{2}X\left( 0\right) .
\end{eqnarray*}

In the pseudospin langage, a bias acts as a pseudo-magnetic field that
couples to the $z$ component of the total pseudospin while the tunneling
acts as a pseudo-magnetic field that couples to the $x$ component of the
total pseudospin. The positive sign in front of the first term on the r.h.s
of Eq. (\ref{enerpseudos}) is due to our particular choice of mapping ($%
R\rightarrow +$ and $L\rightarrow -$) for the pseudospin states. A positive
bias forces the $z$ component of the pseudospin down; i.e. pushes the
electronic charge in the left well. At zero bias, there is equal population
of electrons in both wells and $\left\langle P_{z}\left( \mathbf{0}\right)
\right\rangle =0$ in order to minimize the capacitive energy.

\subsection{The coherent liquid state at $\protect\nu =1$}

At $\nu =1$, the 2DEG can have spontaneous interlayer coherence. For nonzero
Zeeman coupling, the ground state is well described (but this description is
only exact at $d=0$) by a state where all electrons are in the symmetric
combination of both wells and all spins are polarized. The order parameters
are then 
\begin{eqnarray*}
\left\langle \rho _{1,1}\left( 0\right) \right\rangle &=&\left\langle \rho
_{3,3}\left( 0\right) \right\rangle =\frac{1}{2}, \\
\left\langle \rho _{1,3}\left( 0\right) \right\rangle &=&\frac{1}{2}%
e^{i\theta },
\end{eqnarray*}%
irrespective of the value of $d$ (for $t\neq 0$, $\theta =0$ where $\theta $
is the angle between the pseudospin vector and the $x$ axis). In the absence
of tunneling, the coherent liquid phase supports a gapless pseudospin wave
excitation\cite{fertigdispersion} that disperses linearly with $q$ (for $%
d\neq 0$) at small wavectors and becomes soft at an interlayer separation $%
d_{c}/\ell \approx 1.2$. This critical separation is increased by a finite
tunneling. Above this critical interlayer separation, the interwell
coherence is lost. The system is then believed to be formed of two composite
fermions liquids with filling factor $\nu _{R}=\nu _{L}=1/2.$ This state is
not captured by the HFA which instead predicts a transition to a
charge-density-wave state.

\subsection{Influence of a bias}

With a bias, the symmetric and antisymmetric states of the non-interacting
2DEG are replaced by the bonding and anti-bonding states defined by%
\begin{eqnarray*}
\left\vert B\right\rangle &=&\sqrt{\frac{1-\sigma }{2}}\left\vert
R\right\rangle +\sqrt{\frac{1+\sigma }{2}}\left\vert L\right\rangle , \\
\left\vert AB\right\rangle &=&\sqrt{\frac{1+\sigma }{2}}\left\vert
R\right\rangle -\sqrt{\frac{1-\sigma }{2}}\left\vert L\right\rangle ,
\end{eqnarray*}%
\newline
where $\sigma =\frac{\Delta _{b}}{\sqrt{\Delta _{b}^{2}+\Delta _{SAS}^{2}}}$
is the unbalance parameter. At $\nu =1$, the ground state has all electrons
in the \bigskip $\left\vert B\right\rangle $ state (i.e. the symmetric state
in this case) with up spin if the Zeeman coupling is non zero.

When interactions are included, we can still easily solve Eq. (\ref{motion})
in the presence of an electric bias, at $\nu =1$, assuming that the Zeeman
term is non zero so that the 2DEG remains spin polarized. We find%
\begin{eqnarray*}
\left\langle \rho _{1,1}\left( 0\right) \right\rangle &=&\nu _{R}=\frac{1}{2}%
\left( 1-\sigma \right) , \\
\left\langle \rho _{3,3}\left( 0\right) \right\rangle &=&\nu _{L}=\frac{1}{2}%
\left( 1+\sigma \right) , \\
\left\langle \rho _{1,3}\left( 0\right) \right\rangle &=&\alpha =\frac{1}{2}%
e^{i\theta }\sqrt{1-\sigma ^{2}}.
\end{eqnarray*}%
When $t=0$, the energy of the 2DEG is again invariant with respect to $%
\theta $ while for $t\neq 0$, the energy is minimized when $\theta =0$. The
bias acts as a pseudomagnetic field that forces the pseudospin up or down
from the $xy$ plane. The interlayer coherence is maintained but diminished.
In the HFA, there is a critical interlayer separation $d_{c}^{HF}\left(
\Delta _{b}\right) $ where interlayer coherence is lost and all the charge
is transferred into one well. Notice that the pseudospin mode remains
gapless under bias although it now becomes soft at a critical interlayer
separation $d_{c}^{GRPA}\left( \Delta _{b}\right) <$ $d_{c}^{HF}\left(
\Delta _{b}\right) $ that depends on bias. This situation is represented in
Fig. 1 in the case of zero tunneling (essentially the same calculation can
be done at nonzero tunneling. As expected, the critical interlayer
separation increases with $\Delta _{SAS}$). For a 2DEG initially in the
incoherent state at $d>d_{c}^{GRPA}\left( \Delta _{b}\right) $, it is
possible to get a coherent state (and so a quantum Hall effect) by
increasing the bias. This transition has been studied in detail both
theoretically\cite{biastheory} and experimentally\cite{biasexperiment}.

\begin{figure}[tbph]
\includegraphics[scale=1]{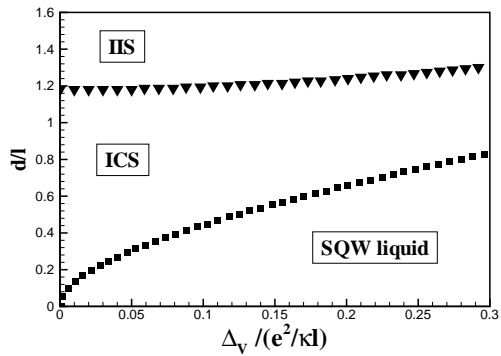}
\caption{Phase diagram of the 2DEG in a DQWS at $\protect\nu =1$. The top
critical line (triangles) is obtained by the instability of the pseudospin
wave mode in the GRPA. As $d/\ell $ goes through this line, the interlayer
coherent state (ICS) loses its coherence to become two interlayer incoherent
states (IIS) with $\protect\nu _{R\left( L\right) }=1/2$. When bias is
increased at fixed $d/\ell $, the ICS looses its coherence and all the
charge is transferred into a single quantum well (SQW liquid). This line
(squares) is obtained in the HFA.}
\label{fig1}
\end{figure}

\section{Topological excitations in the DQWS}

\subsection{Spin and pseudospin skyrmions}

In a monolayer system with only spin degrees of freedom, Eq. (\ref%
{enerpseudos}) becomes

\begin{eqnarray}
\frac{E_{HF}}{N} &=&-\frac{\Delta _{Z}}{\nu }\left\langle S_{z}\left( 
\mathbf{0}\right) \right\rangle  \label{eskyrmion} \\
&&+\frac{1}{4\nu }\sum_{\mathbf{G}}\Upsilon _{2}\left( \mathbf{G}\right)
\left\vert \left\langle \rho \left( \mathbf{G}\right) \right\rangle
\right\vert ^{2}  \notag \\
&&-\frac{1}{\nu }\sum_{\mathbf{G}}X\left( \mathbf{G}\right) \left\vert
\left\langle \mathbf{S}\left( \mathbf{G}\right) \right\rangle \right\vert
^{2},  \notag
\end{eqnarray}%
where the interaction%
\begin{equation*}
\Upsilon _{2}\left( \mathbf{G}\right) =2H\left( \mathbf{G}\right) -X\left( 
\mathbf{G}\right) .
\end{equation*}

In the absence of Coulomb electrostatic energy and Zeeman coupling and in
the gradient approximation where $\left\langle \mathbf{S}\left( \mathbf{r}%
\right) \right\rangle $ is assumed to vary smoothly over the magnetic length 
$\ell $, the energy in Eq. (\ref{eskyrmion}) can be reduced to that of the
O(3) nonlinear sigma model.\cite{macdobible} The O(3) nonlinear sigma model
in a planar geometry possesses topological solitons or skyrmions. Using the
complex function $w\left( z\right) $ which represents the stereographic
projection of the unit sphere of spin textures $\mathbf{s}$, a skyrmion of
Pontryagin index $Q=1$ (which corresponds to the addition of one electron to
the 2DEG) can be written\cite{ezawalivre} as%
\begin{equation}
w\left( z\right) =\frac{s_{x}\left( z\right) -is_{y}\left( z\right) }{%
1-s_{z}\left( z\right) }=\frac{z-b}{\lambda },  \label{wz}
\end{equation}%
where $z=x+iy.$ Eq. (\ref{wz}) describes a skyrmion of size $\left\vert
\lambda \right\vert $ centered at position $b=x_{b}+iy_{b}$ in the $x-y$
plane. The spin components are

\begin{eqnarray*}
s_{x}\left( z\right) -is_{y}\left( z\right) &=&\frac{2\lambda ^{\ast }\left(
z-b\right) }{\left\vert \lambda \right\vert ^{2}+\left\vert z-b\right\vert
^{2}}, \\
s_{z}\left( z\right) &=&\frac{\left\vert z-b\right\vert ^{2}-\left\vert
\lambda \right\vert ^{2}}{\left\vert z-b\right\vert ^{2}+\left\vert \lambda
\right\vert ^{2}}.
\end{eqnarray*}%
At infinity, the spins point upward while at the center $z=b$ of the
skyrmion, they point downward. The phase $\varphi $ in $\lambda =\left\vert
\lambda \right\vert e^{i\varphi }$ fixes the global orientation of the spins
forming the skyrmion.

In quantum Hall systems, skyrmion-antiskyrmion excitations have been shown
to have lower energy than the corresponding maximally spin-polarized
Hartree-Fock electron-hole excitation if the Zeeman coupling is not too
strong.\cite{fertigskyrmion,sondhi} Around filling factor $\nu =1$, a finite
density of skyrmions ($\nu >1$) or antiskyrmions ($\nu <1$) are included in
the groundstate (with filling fraction $\nu _{s}=\left\vert \nu
-1\right\vert $). At $T=0$ K, these quasiparticles condense into a Skyrme
crystal.

The equation of motion method derived in Ref. \onlinecite{cotemethode} was
used some years ago to study the phase diagram of the Skyrme crystal in the $%
\Delta _{Z}-\nu _{s}$ space. Numerical results\cite{breyprl,cotegirvinprl}
show that, in a large portion of the phase space, skyrmions crystallize into
a square lattice structure with two skyrmions of opposite phases ($\varphi
=0 $ and $\varphi =\pi $) per unit cell. This arrangement was termed $SLA$
(for square lattice antiferromagnetic). More recent calculations\cite%
{cotegroup1} show that, as the Zeeman coupling is increased from small
values at fixed quasiparticle filling $\nu _{s}$, the crystal structure
changes from a checkerboard\cite{breymeron} lattice of merons near $\Delta
_{Z}=0$ (i.e. a $SLA$ skyrmion lattice where each skyrmion splits into two
merons of charge $e/2$ with opposite vorticities and where all merons are
equally spaced) to a lattice of biskyrmions\cite{nazarov} at small $\Delta
_{Z},$ a $SLA\ $skyrmion lattice at moderate $\Delta _{Z}$ and finally into
a $TL120$ lattice of skyrmions (a triangular lattice with three skyrmions
with phases $\varphi =0,2\pi /3,4\pi /3$ per unit cell to avoid the
frustration created by the preferred antiferromagnetic ordering of the
skyrmions\cite{nazarov}) at higher values of $\Delta _{Z}$. The spin texture
is gradually lost as the Zeeman coupling further increases and we finally
have a Wigner crystal of maximally polarized quasiparticles with no spin
texture.

For a spin-polarized 2DEG in a DQWS, the energy functional of Eq. (\ref%
{enerpseudos}) becomes 
\begin{eqnarray}
\frac{E_{HF}}{N} &=&\frac{\Delta _{b}}{\nu }\left\langle P_{z}\left( \mathbf{%
0}\right) \right\rangle -\frac{\Delta _{SAS}}{\nu }\left\langle P_{x}\left( 
\mathbf{0}\right) \right\rangle  \notag  \label{edouble} \\
&&+\frac{1}{4\nu }\sum_{\mathbf{G}}\Upsilon _{2}\left( \mathbf{G}\right)
\left\vert \left\langle \rho \left( \mathbf{G}\right) \right\rangle
\right\vert ^{2}  \label{edoublep} \\
&&+\frac{1}{\nu }\sum_{\mathbf{G}}J_{z,2}\left( \mathbf{G}\right) \left\vert
\left\langle P_{z}\left( \mathbf{G}\right) \right\rangle \right\vert ^{2} 
\notag \\
&&-\frac{1}{\nu }\sum_{\mathbf{G}}\sum_{\alpha =+,-}\widetilde{X}\left( 
\mathbf{G}\right) \left\vert \left\langle \mathbf{P}_{\bot ,\alpha }\left( 
\mathbf{G}\right) \right\rangle \right\vert ^{2},  \notag
\end{eqnarray}%
with 
\begin{equation*}
J_{z,2}\left( \mathbf{G}\right) =H\left( \mathbf{G}\right) -\widetilde{H}%
\left( \mathbf{G}\right) -X\left( \mathbf{G}\right) .
\end{equation*}%
In the absence of bias, tunneling and Coulomb electrostatic energies and in
the gradient approximation, the energy in Eq. (\ref{edoublep}) can be
reduced to that of the anisotropic nonlinear sigma model with a unit
pseudospin field $\mathbf{p}\left( \mathbf{r}\right) $.\cite{macdobible} The
topological excitations of this model are bimerons and merons (or
pseudospin-skyrmions and pseudospin-merons). A bimeron with topological
charge $Q=1$ has its pseudospin field given by\cite{ezawalivre}%
\begin{eqnarray*}
p_{x}\left( z\right) -ip_{y}\left( z\right) &=&\frac{2\left( z-b\right)
\left( z^{\ast }+b^{\ast }\right) }{\left\vert z-b\right\vert
^{2}+\left\vert z+b\right\vert ^{2}}, \\
p_{z}\left( z\right) &=&\frac{\left\vert z-b\right\vert ^{2}-\left\vert
z+b\right\vert ^{2}}{\left\vert z-b\right\vert ^{2}+\left\vert
z+b\right\vert ^{2}},
\end{eqnarray*}%
where $\pm b$ are the positions of the two merons forming the bimeron.
Alternatively, we can write 
\begin{eqnarray*}
w\left( z\right) &=&\frac{p_{x}\left( z\right) -ip_{y}\left( z\right) }{%
1-p_{z}\left( z\right) } \\
&=&\left( \frac{z-z_{L}}{z-z_{R}}\right) e^{-i\varphi } \\
&=&\left( \frac{z-b}{z+b}\right) e^{-i\varphi },
\end{eqnarray*}%
where $z_{R\left( L\right) }$ are the positions of the merons in the right
and left wells. The angle $\varphi $ gives the global orientation of the
pseudospin vectors with respect to the $x$ axis at infinity. When $z=b(-b)$,
we are at the center of the meron on the left(right) well and there the
pseudospin $p_{z}=-1(+1)$.

Bimerons and merons have been studied extensively in the context of the QHE.%
\cite{ezawabimeron,rajaramanbimeron,breycristalbimerons} Again,
bimeron-antibimeron excitations have been shown to be the relevant
excitations near $\nu =1.$ Although we have not performed an exhaustive
calculation of the phase diagram of bimeron crystals, our numerical results%
\cite{cotegroup1} show that at finite tunneling, an $SLA$ (or rectangular
antiferromagnetic) configuration of bimerons is stable up to an interlayer
separation $d/\ell \approx 1$. At very small tunneling, the bimeron lattice
becomes a lattice of merons with again the checkerboard configuration. In
comparison with Wigner crystal in bilayer systems where a one-component
Wigner crystal can only be stabilized at small interlayer distances of order 
$d/\ell \approx 0.1$, the interlayer coherence in a bimeron or meron lattice
persists to much larger $d/\ell .$

\subsection{CP$^{3}$ skyrmion}

When both spin and pseudospin are active degrees of freedom, electronic
states must be described by a four-component spinor. As explained in Ref. %
\onlinecite{rajaramancp3}, this spinor is a CP$^{3}$ spinor since the DQWS
has a U(1) gauge invariance (all four components of the spinor must be
transformed by the same phase for the DQWS's energy to remain invariant).
Strictly speaking, a texture of a CP$^{3}$ spinor can lie wholly in the spin
degrees of freedom, or wholly in the pseudospin degrees of freedom. We only
have a guarantee that the topological charge associated with the texture
integrates to an integer. In this paper we use the phrase \textquotedblleft
CP$^{3}$ Skyrmion\textquotedblright\ to refer to textures in which \textit{%
both} the spin and pseudospin degrees of freedom are appreciably varying.
This is sometimes referred to as an \textit{interwined texture}.

To start the iteration process needed to solve Eq. (\ref{motion}), we must
provide an approximate solution for the crystal of CP$^{3}$ skyrmions. From
our discussion above, we expect that an $SLA$ configuration of skyrmions
could be a likely solution. Following Rajaraman\cite{rajaramancp3}, we write
the four-component spinor%
\begin{equation}
\Psi \left( \mathbf{r}\right) =A\left( 
\begin{array}{c}
z-b \\ 
\lambda _{1} \\ 
z+b \\ 
\lambda _{2}%
\end{array}%
\right) ,  \label{solutioncp3}
\end{equation}%
were $z=x+iy$ and the normalisation factor is 
\begin{equation*}
A=\frac{1}{\sqrt{\left\vert \lambda _{1}\right\vert ^{2}+\left\vert \lambda
_{2}\right\vert ^{2}+2\left\vert z\right\vert ^{2}+\left\vert b\right\vert
^{2}}}.
\end{equation*}%
This state contains a skyrmion of size $\left\vert \lambda _{1}\right\vert $
at position $b$ in the right well, a skyrmion of size $\left\vert \lambda
_{2}\right\vert $ at position $-b$ in the left well and a bimeron in the
spin up component of the pseudospin centered at $z=0$. The CP$^{3}$ static
energy of this skyrmion is given by%
\begin{equation}
E_{CP^{3}}=2\rho _{s}\sum_{a=1}^{4}\sum_{j=x,y}\int d\mathbf{r}\left(
D_{j}\Psi _{a}\left( \mathbf{r}\right) \right) ^{\ast }\left( D_{j}\Psi
_{a}\left( \mathbf{r}\right) \right) ,  \label{cp3action}
\end{equation}%
where $D_{j}=\partial _{j}-iK_{j}$ with $K$ a gauge field defined by $%
K_{j}=-i\sum_{a}\Psi _{a}^{\ast }\left( \mathbf{r}\right) \partial _{j}\Psi
_{a}\left( \mathbf{r}\right) $ and $\rho _{s}$ a \textquotedblleft
spin-pseudospin\textquotedblright\ stiffness. The field $\Psi \left( \mathbf{%
r}\right) $ must satisfy the constraint $\sum_{a}\left\vert \Psi _{a}\left( 
\mathbf{r}\right) \right\vert ^{2}=1$. Eq. (\ref{cp3action}) can be obtained
from our Hartree-Fock energy by taking the limit $\Delta _{SAS}=\Delta
_{Z}=d=0$. The solution of Eq. (\ref{solutioncp3}) is a skyrmion with
topological charge $Q=\int d\mathbf{r\delta }Q\left( \mathbf{r}\right) =1$.
The CP$^{3}$ topological charge density is defined by%
\begin{equation}
\delta Q\left( \mathbf{r}\right) =-\frac{i}{2\pi }\varepsilon _{ij}\left(
D_{i}\Psi _{a}\left( \mathbf{r}\right) \right) ^{\ast }\left( D_{j}\Psi
_{a}\left( \mathbf{r}\right) \right) .
\end{equation}

The order parameters for this single quasiparticle state are given, in real
space, by $\left\langle \rho _{i,j}\left( \mathbf{r}\right) \right\rangle
=\left\langle \Psi _{i}^{\dag }\left( \mathbf{r}\right) \Psi _{j}\left( 
\mathbf{r}\right) \right\rangle .$ Fourier-transforming this expression, we
can easily write the $\left\langle \rho _{i,j}\left( \mathbf{q}\right)
\right\rangle ^{\prime }s$ for this state. To write an approximate solution
for a \textit{crystal }of these quasiparticles, we consider the change $%
\delta \left\langle \rho _{i,j}\left( \mathbf{r}\right) \right\rangle $ in
the ground state (at $\nu =1$) when a skyrmion is added to the system 
\begin{equation*}
\delta \left\langle \rho _{i,j}\left( \mathbf{r}\right) \right\rangle
=\left\langle \rho _{i,j}\left( \mathbf{r}\right) \right\rangle
-\left\langle \rho _{i,j}\left( \mathbf{r}\right) \right\rangle _{GS},
\end{equation*}%
where $\left\langle \rho _{i,j}\left( \mathbf{r}\right) \right\rangle _{GS}$
describe the ground state (at $\nu =1$) which has (at zero bias) all
electrons in the bonding (or symmetric) state with up spins (see Sec. II).
In principle, the fields $\delta \left\langle \rho _{i,j}\left( \mathbf{r}%
\right) \right\rangle $ are zero when we are far away from the position of
the skyrmion so that the crystal state can be written approximately as%
\begin{equation*}
\left\langle \rho _{i,j}\left( \mathbf{r}\right) \right\rangle =\left\langle
\rho _{i,j}\left( \mathbf{r}\right) \right\rangle _{GS}+\sum_{\mathbf{R}%
}\sum_{\alpha =1,2}\delta \left\langle \rho _{i,j}^{(\alpha )}\left( \mathbf{%
r-R-c}_{\alpha }\right) \right\rangle ,
\end{equation*}%
where $\mathbf{c}_{\alpha }$ is the position vector of the two skyrmions in
the unit cell, $\mathbf{R}$ is a lattice vector, and $\alpha $ is the index
of the phase of each of the two skyrmions. The order parameters for the
crystal state are then given by%
\begin{equation}
\left\langle \rho _{i,j}\left( \mathbf{G\neq 0}\right) \right\rangle \sim
\sum_{\alpha =1,2}e^{-i\mathbf{G}\cdot \mathbf{c}_{\alpha }}\delta
\left\langle \rho _{i,j}^{(\alpha )}\left( \mathbf{G}\right) \right\rangle .
\label{rogcp3}
\end{equation}%
In general, it is sufficient to give the $\left\langle \rho _{i,j}\left( 
\mathbf{G\neq 0}\right) \right\rangle ^{\prime }s$ given by Eq. (\ref{rogcp3}%
) on the first or first two shells of reciprocal lattice vectors in order
for the program to converge to the CP$^{3}$ skyrmion crystal. The SLA
configuration is obtained by choosing $\mathbf{R}_{1}=na\widehat{\mathbf{x}}$
and $\mathbf{R}_{2}=ma\widehat{\mathbf{y}}$ for the lattice vectors ($%
n,m=0,\pm 1,\pm 2,...$), $\mathbf{c}_{1}=-\frac{a}{4}\widehat{\mathbf{y}}$
and $\mathbf{c}_{2}=+\frac{a}{4}\widehat{\mathbf{y}}$ for the positions of
the two skyrmions. For the first skyrmion, we take $b=-\frac{a}{4}$ and $%
\lambda _{1}=\lambda _{2}=1.$ For the second skyrmion, we take, $b^{\prime }=%
\frac{a}{4}$ and $\lambda _{1}^{\prime }=\lambda _{2}^{\prime }=-1$ in order
to rotate the global phase by $\pi .$

\section{Numerical results on CP$^{3}$ crystals}

\subsection{Crystal states considered}

The HFA does not contain the correlations necessary to describe the ground
state at $d/\ell >1.2$ where interwell coherence is lost. For this reason,
we limit our numerical calculations of crystal states to interlayer
separations $0\leq d/\ell \leq 1.2.$ In the monolayer and polarized bilayer
limits, we found that a square lattice with two skyrmions of opposite phases
per unit cell is the ground state in a wide region of parameter space. We
thus choose to consider the following states in our analysis:

\begin{itemize}
\item CP$^{3}$: a square lattice with two spin-pseudospin skyrmions of
opposite phases per unit cell as described at the end of Sec. III. This
state is represented in Fig. 2 in the case of small tunneling where each
skyrmion is broken into two merons of opposite vorticities.

\item SPB: a spin-polarized square lattice with two bimerons of opposite
phases per unit cell. The spinor of Eq. (\ref{solutioncp3}) is replaced by $%
\Psi \left( \mathbf{r}\right) =A\left( 
\begin{array}{c}
z-b_{1} \\ 
0 \\ 
z+b_{1} \\ 
0%
\end{array}%
\right) .$ At small tunneling, the bimerons split into a pair of two merons
with opposite vorticities. This spin-polarized bimeron (or merons) lattice
should be the ground state when the Zeeman energy is of the order or bigger
than the tunneling energy.

\item SS: this is a symmetric skyrmion state which is a pseudospin-polarized
square lattice state with two symmetric-band spin-skyrmions of opposite
phases per unit cell. By \textquotedblleft symmetric-band\textquotedblright
, we mean that the SU(2)\ spinor for the first electron would be given by $%
\Psi \left( \mathbf{r}\right) =A^{\prime }\left( 
\begin{array}{c}
z \\ 
\lambda _{1} \\ 
0 \\ 
0%
\end{array}%
\right) $ in the symmetric-antisymmetric basis $(S+;$ $S-;AS+;AS-$) basis or
by $\Psi \left( \mathbf{r}\right) =A\left( 
\begin{array}{c}
z \\ 
\lambda _{1} \\ 
z \\ 
\lambda _{1}%
\end{array}%
\right) $ in the $(R+;$ $R-;L+;L-$) basis. We expect this phase to be the
ground state state when tunneling energy dominates the Zeeman energy.

\item High Tunneling CP$^{3}$(HCP$^{3}$): a square lattice with two
spin-pseudospin skyrmion of opposite phase per unit cell. The difference
between this state and the CP$^{3}$ state above is that here the spin
texture exists in the symmetric and antisymmetric bands while in the latter
it exists separately in each quantum well. This state is the ground state
only when the tunneling energy is higher than the Zeeman energy and only for
filling factor $\nu >1$. We note that the HCP$^{3}$ state is an intermediate
state between SS and SPB states. The textures in the HCP$^{3}$ state splits
into two vortices with charge $e/2$ by reducing the Zeeman gap.
\end{itemize}

When the tunneling or Zeeman couplings increase, the size of the pseudospin
or spin skyrmions decreases. At large values of both these parameters, a
limit is reached where the skyrmion state reaches maximal spin and
pseudospin polarization. The resulting state may be viewed as a crystal of
Hartree-Fock quasiparticles. When interlayer coherence is non zero, the HF\
quasiparticles are delocalized in both wells and form a coherent Wigner
crystal (i.e. a \textquotedblleft one-component \textquotedblright Wigner
crystal)

\begin{figure}[tbph]
\includegraphics[scale=0.18]{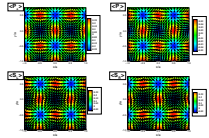}
\caption{(Color online). Spin textures in each well (bottom) and pseudospin textures in each
spin component (top) of a CP$^{3}$ crystal for $\protect\nu =0.8,\Delta
_{SAS}/\left( e^{2}/\protect\kappa \ell \right) =0.0002,$ $\Delta
_{Z}/\left( e^{2}/\protect\kappa \ell \right) =0.002,d/\ell =1.0$. The
crystal has four merons per unit cell. In each unit cell, two merons with
the same vorticities have opposite phases as explained in the text. The
contour color indicated on the legends at the right side of each plot is for
the $z$ component of each field.}
\label{figure2}
\end{figure}

With increasing bias, the CP$^{3}$ or SS solution will continuously
transform into a monolayer spin skyrmion. The SS solution, for example can
be written in the $R/L$ basis as $\Psi \left( \mathbf{r}\right) =\frac{A}{%
\sqrt{2}}\left( 
\begin{array}{c}
\sqrt{1-\sigma }z \\ 
\sqrt{1-\sigma }\lambda _{1} \\ 
\sqrt{1+\sigma }z \\ 
\sqrt{1+\sigma }\lambda _{1}%
\end{array}%
\right) $ in the presence of a bias (where $\sigma $ is the unbalance
parameter and $\sigma =0$ in the absence of bias).

\subsection{Effect of interlayer separation at zero bias}

We first compute the energy of the three states just introduced as a
function of the interlayer separation for parameters $\nu =0.8,$ $\Delta
_{SAS}/\left( e^{2}/\kappa \ell \right) =0.0002$ and $\Delta _{Z}/\left(
e^{2}/\kappa \ell \right) =0.002$. Figure 3(a) shows the energy differences $%
E_{CP^{3}}-E_{SPB}$ and $E_{SS}-E_{SPB}$. At small interlayer separation,
the ground state is the SPB crystal while above some critical interwell
separation that depends on the Zeeman coupling, a CP$^{3}$ crystal state
emerges and remains the ground state up to the largest value of $d/\ell $
that we consider \textit{(i.e.} $d/\ell =1.2$). Figure 3(b) shows the
difference in energy between the CP$^{3}$ and the SPB crystals for several
values of the Zeeman coupling. As expected, the interlayer separation at
which the SPB-CP$^{3}$ transition takes place increases with increasing
Zeeman coupling. (We remark that $e^{2}/\kappa \ell =50.\,\allowbreak 489%
\sqrt{B}$ K, so that the difference in energy, at $d/\ell =1.0$ and $\Delta
_{Z}/\left( e^{2}/\kappa \ell \right) =0.002$, is of the order of $90$ mK.)
Moreover, in Fig. 3(c), we see that the spin polarization per electron $%
\left\langle S_{z}\left( \mathbf{q}=0\right) \right\rangle /\nu $ varies
strongly with interlayer separation in the CP$^{3}$ crystal state in
comparison with the SS state. The spin depolarization of the CP$^{3}$ state
increases with decreasing $\Delta _{Z}$ and reaches a maximum at about $%
d/\ell \approx 1$. Figure 3(d) shows the behaviour of the pseudospin
polarization per electron in the $x-y$ plane i.e. $\left\langle P_{x}\left( 
\mathbf{q}=0\right) \right\rangle /\nu .$ The pseudospin polarization
increases when the spin polarization decreases. The value of pseudospin
polarization gives some indication of the size of the bimeron in the CP$^{3}$
skyrmions. When $\left\langle P_{x}\left( \mathbf{q}=0\right) \right\rangle
/\nu =0.5$, there are no pseudospin vortices in the phase considered. Note
that we have chosen in our analysis a very small value of the tunneling
constant. Our results of this section stay essentially the same if the
tunneling coupling is exactly zero.

The observation that this CP$^{3}$ state is most prominent at large $d$
suggests that it is stabilized by the interlayer charging energy. The merons
of the SPB state involve \textquotedblleft tilting\textquotedblright\ of the
pseudospin at their cores into one layer or the other, at an energy cost of
order $e^{2}d/\kappa \ell ^{2}$. For large enough $d$, it is energetically
favorable to admix spin states so that near their cores the charge of the
vortices will be balanced. An examination of the charge densities in each
well reveals that the CP$^{3}$ lattice is indeed more locally balanced than
the SPB lattice.

\subsection{Effect of tunneling at zero bias}

In Fig. 4, we show the behavior of the spin polarization per electron with
filling factor for $\nu <1$ for two values of the interlayer separation. At $%
\nu =1$, $\left\langle S_{z}\left( \mathbf{q}=0\right) \right\rangle /\nu
=0.5$. Away from $\nu =1$, the spin polarization decreases rapidly for the CP%
$^{3}$ crystal until it reaches a minimum at about $\nu =0.9$. Then, as $\nu 
$ is further decreased and the density of skyrmions increases, the size of
the skyrmions also decreases and the Wigner crystal limit is reached where
the ground state is again fully spin polarized. The behavior of the spin
polarization we find for the CP$^{3}$ crystal is identical to what was found
for Skyrme crystals in a single layer system.\cite{breyprl} As $\nu $ is
increased towards $\nu =1$, we also find that the critical interlayer
separation for the transition between the SPB and the CP$^{3}$ states
decreases so that the CP$^{3}$ crystal state is stable over a larger range
of interlayer separation for smaller quasiparticle filling. For the SS state
that occurs at high tunneling, the variation of the spin polarization is
less marked than in the CP$^{3}$ crystal.

\begin{figure}[tbph]
\includegraphics[scale=1]{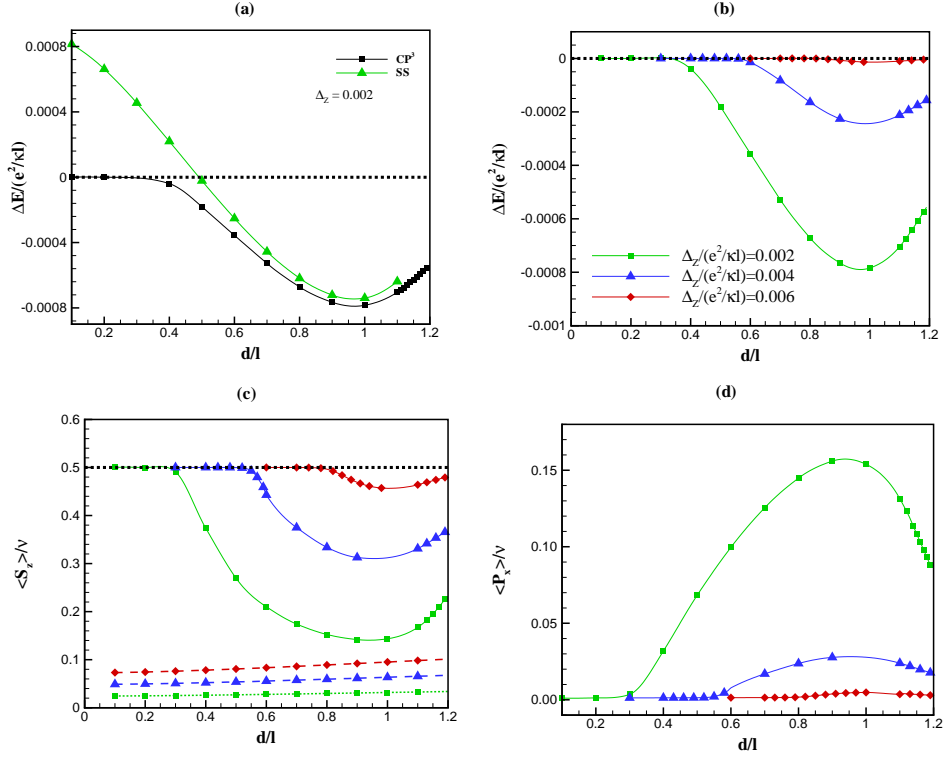}
\caption{(Color online). CP$^{3}$ crystal at $\protect\nu =0.8$ and $\Delta
_{SAS}=0.0002\left( e^{2}/\protect\kappa \ell \right) $. (a) Energy
difference between the CP$^{3}$ or SS state and the SPB\ state at $\Delta
_{Z}=0.002\left( e^{2}/\protect\kappa \ell \right) $; (b) energy difference
between the CP$^{3}$ and SPB state for different values of the Zeeman
coupling; (c) Spin polarization and (d) pseudospin polarization per electron
in the CP$^{3}$ state (full lines) and SS state (dashed lines) for the
different Zeeman couplings indicated in (b).}
\label{fig3}
\end{figure}

\begin{figure}[tbph]
\includegraphics[scale=1]{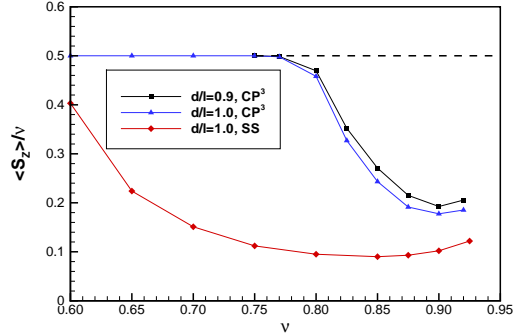}
\caption{(Color online). Spin polarization in the CP$^{3}$ and SS skyrmion
crystals as a function of filling factor for $\protect\nu <1$. Here $\Delta
_{Z}/\left( e^{2}/\protect\kappa \ell \right) =0.006$. For the CP$^{3}$
curves, $\Delta _{SAS}/\left( e^{2}/\protect\kappa \ell \right) =0.0002,$%
while for the SS curve, $\Delta _{SAS}/\left( e^{2}/\protect\kappa \ell
\right) =0.05.$}
\label{figure4}
\end{figure}

Because the spin polarization is minimal around $d/\ell =1$ for $\nu =0.8,$
we choose this value of the interlayer separation to study the effect of
tunneling on the spin polarization. Figure 5 shows the difference in energy $%
E_{CP^{3}}-E_{SPB}$ and $E_{SS}-E_{SPB}$ for three values of the Zeeman
coupling. At small Zeeman coupling, increasing $\Delta _{SAS}$ causes a
transition from the CP$^{3}$ to the SS crystal. At larger Zeeman couplings,
where the ground state is the SPB crystal at zero tunneling, increasing $%
\Delta _{SAS}$ causes first a transition to a CP$^{3}$ crystal (in a very
narrow range of $\Delta _{SAS}$) and then into a SS at larger tunneling
values. The CP$^{3}$ crystal exists only in narrow range of $\Delta _{SAS}$
and that range decreases with increasing Zeeman coupling so that the CP$^{3}$
state disappears at large $\Delta _{Z}$. Typically, the SS phase occurs for $%
\Delta _{SAS}\gtrsim \Delta _{Z}/2$. 
\begin{figure}[tbph]
\includegraphics[scale=1]{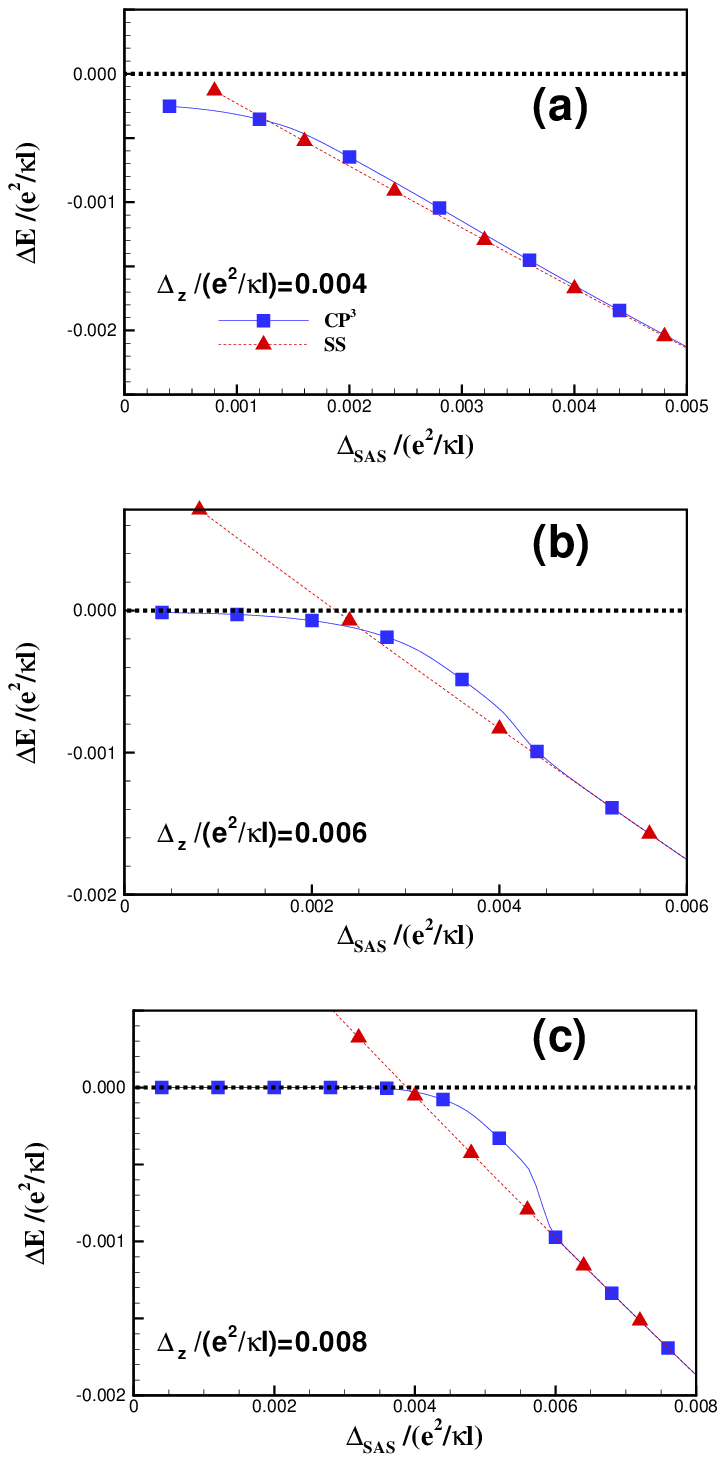}
\caption{(Color online). Energy difference between the CP$^{3}$ crystal
(filled squares) or SS crystal (filled triangles) and the SPB state at $%
\protect\nu =0.8$ and $d/\ell =1.0$ as a function of tunneling for Zeeman
couplings: (a) $\Delta _{Z}/\left( e^{2}/\protect\kappa \ell \right) =0.004;$%
(b) $\Delta _{Z}/\left( e^{2}/\protect\kappa \ell \right) =0.006;$ and (c) $%
\Delta _{Z}/\left( e^{2}/\protect\kappa \ell \right) =0.008.$}
\label{figure5}
\end{figure}

The change in the spin and pseudospin polarizations of the CP$^{3}$ and SS
crystals with $\Delta _{SAS}$ is shown in Fig. 6 for the values of the
Zeeman coupling considered in Fig. 5. We see from this figure that
increasing $\Delta _{SAS}$ increases the pseudospin polarization and
decreases the spin polarization. At the transition from the CP$^{3}$ to the
SS state indicated by the vertical dashed lines in Fig. 6, there is a sharp
drop in the spin polarization. This sudden change in $\left\langle
S_{z}\left( \mathbf{q}=0\right) \right\rangle ,$ and so in the in-plane spin
polarization, should lead to abrupt changes in the NMR\ relaxation time.

\begin{figure}[tbph]
\includegraphics[scale=1]{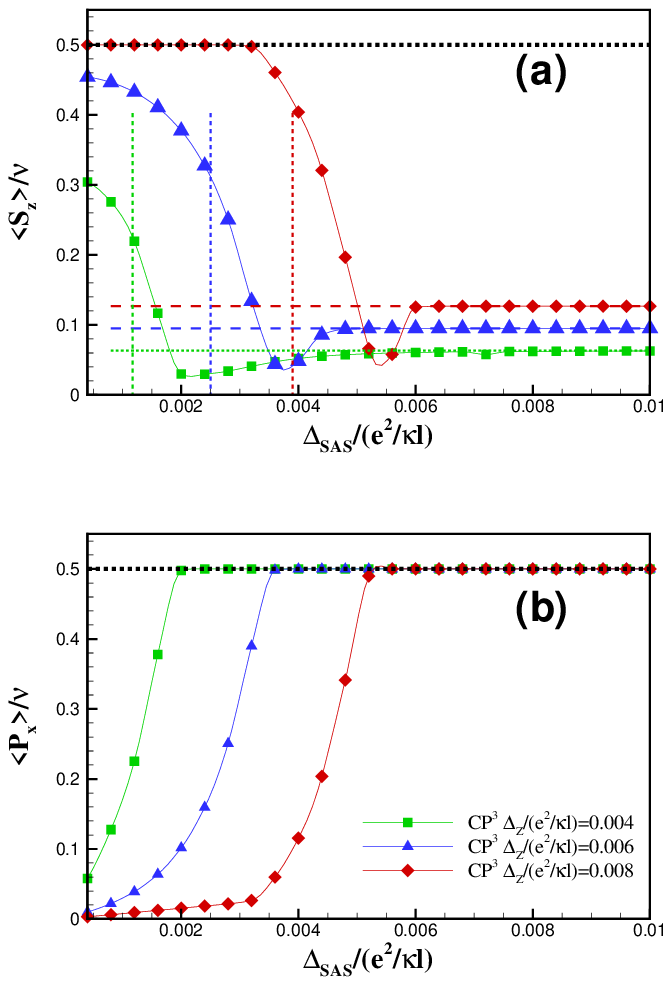}
\caption{(Color online). Spin (a) and pseudospin (b) polarization per
electron as a function of tunneling for the CP$^{3}$ crystal (lines with
symbols) and the SS state (dashed lines) for different values of the Zeeman
coupling. In (a), the spin polarization goes to that of the SS state at
large tunneling. The vertical lines in (a) indicate the critical tunneling
for the transition between the CP$^{3}$ and the SS states as found from Fig.
5.}
\label{figure6}
\end{figure}

Our results, so far, have been for filling factor $\nu <1$. Interestingly,
we do not find a precisely analogous spin-pseudospin configuration for
filling factors $\nu >1$. Instead, as explained in Section IV, we have found
an intermediate spin-pseudospin state at large values of $\Delta _{SAS}$ and
small separations ( $d/\ell \lesssim 0.7$ ), which we call the HCP$^{3}$
state. Fig. 7(a) shows the difference in energy of HCP$^{3}$ or SS and SPB.
As we can see in this figure, by increasing $\Delta _{SAS}$ the ground state
changes from SPB to HCP$^{3}$ and then to SS. Also in Fig. 7(b) we can see
the spin depolarization in HCP$^{3}$ state as a function of $\Delta _{SAS}$.
The spin polarization of the HCP$^{3}$ state is more sensitive to $\Delta
_{SAS}$ than the SS state. 
\begin{figure}[tbph]
\includegraphics[scale=1]{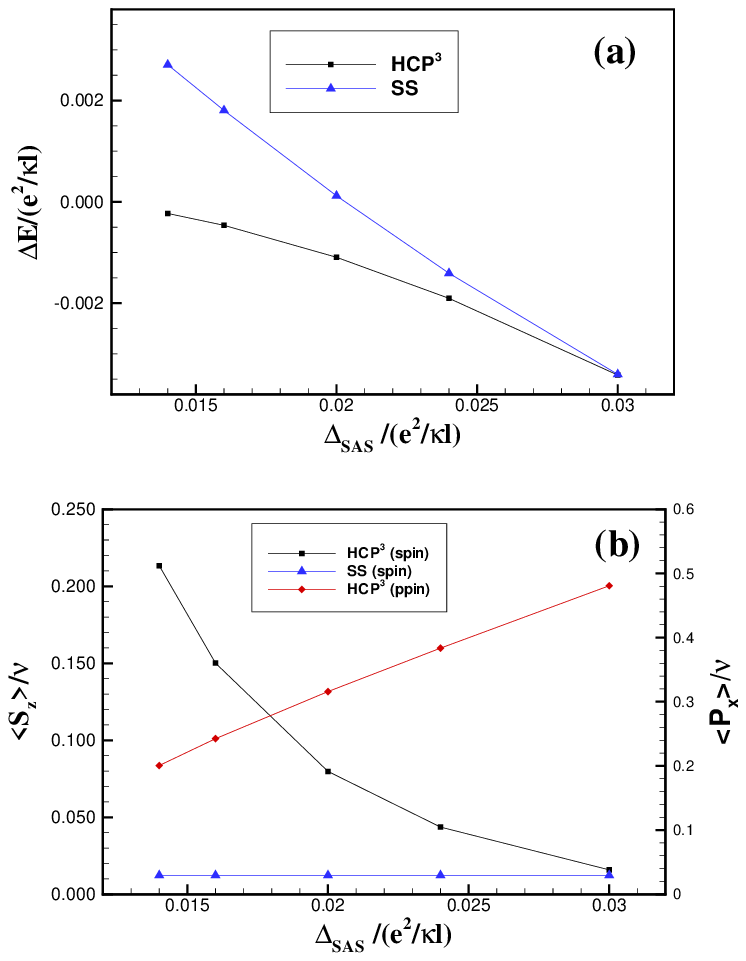}
\caption{(Color online). (a) Energy difference between the HCP$^{3}$ crystal
or SS crystals and the SPB state as a function of tunneling for $\protect\nu %
=1.2,\Delta _{Z}/\left( e^{2}/\protect\kappa \ell \right) =0.0015$ and $%
d/\ell =0.1$; (b) Spin and pseudospin polarization as a function of
tunneling for the HCP$^{3}$ and SS crystals.}
\label{figure7}
\end{figure}

For $\nu >1$ the SS state also can be the ground state. At $\nu =1.04,$ we
find that the SS state is the ground state for $d/\ell \lesssim 0.8$, $%
\Delta _{SAS}/\left( e^{2}/\kappa \ell \right) >.03$ and $0<\Delta
_{Z}/\left( e^{2}/\kappa \ell \right) \lesssim 0.002.$ The spin polarization
with interlayer separation in the SS state is shown in Fig. 8. The linear
behaviour is typical of what is obtained for $\nu <1$ in the SS state (see
Fig. 3, for example).

\begin{figure}[tbph]
\includegraphics[scale=1]{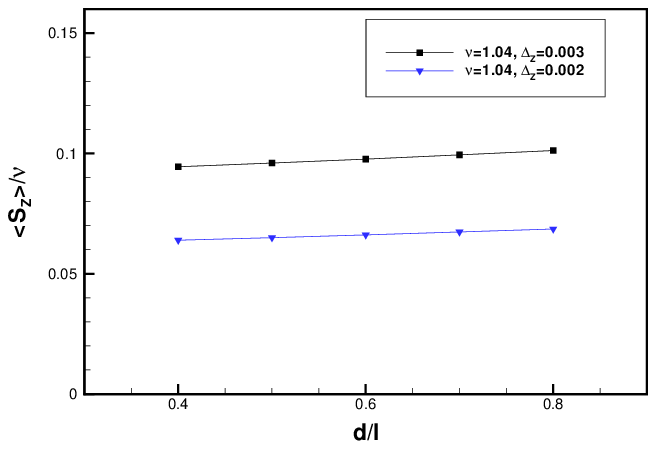}
\caption{(Color online). Spin polarization as a function of interlayer
separation in the SS state for different values of the Zeeman coupling and
filling factors. Here $\Delta _{SAS}/\left( e^{2}/\protect\kappa \ell
\right) =0.04$ and $\protect\nu =1.04.$}
\label{figure8}
\end{figure}

\subsection{Effect of a bias on the spin polarization}

To conclude this section, we look at the effect of a potential bias on the
spin polarization. Intuitively we understand that a CP$^{3}$ skyrmion
involves a \textquotedblleft twist\textquotedblright\ in some high
dimensional space that is difficult to plot in 2D. That twist will occur
through degrees of freedom where it costs the least energy, and the texture
will vary slowly in sectors where the system is \textquotedblleft
stiff\textquotedblright . If we can change the stiffness of textures along
some direction of phase space we can drive the texture into or out of that
degree of freedom. A simple analogy would be to drive an O(3) model into an
XY model by making excursions into the z-direction too costly.

The behavior of this system with respect to bias illustrates this physics.
We choose the parameters $d/\ell =1,~\nu =0.8,~\Delta _{SAS}/\left(
e^{2}/\kappa \ell \right) =0.0002$ and $\Delta _{Z}/\left( e^{2}/\kappa \ell
\right) =0.01$ so that, at the balanced point, the ground state is a
spin-polarized meron crystal (SPB). Our numerical results, plotted in Fig.
9, show that there is a transition first into a CP$^{3}$ crystal and then
into the SS state as the applied bias increases. The energy of the CP$^{3}$
crystal interpolates nicely between the SPB and SS phases as can be seen in
the figure. The corresponding spin polarizations are shown in Fig. 9(b). The
bias has the effect of inducing a linear spin depolarization of the 2DEG in
the CP$^{3}$ state. In effect, the texture inducing the deviation of charge
density from $\nu =1$ is being shifted from the pseudospin degree of freedom
to the spin degree of freedom in a continuous way. Note that the spin
polarization varies only slightly with $d/\ell $ in the SS state. Fig. 9(c)
shows the filling factor $\nu _{R}$ and $\nu _{L}$ in the CP$^{3}$ state
(the exact same curves are obtained in the SS state). Above $\Delta
_{b}/\left( e^{2}/\kappa \ell \right) \approx 0.30,$ all the charge is
transferred into the right layer and the spin polarization is that
appropriate for a monolayer Skyrmion crystal with filling factor $\nu =0.8$
and Zeeman coupling $\Delta _{Z}/\left( e^{2}/\kappa \ell \right) =0.01$ and
is then independent of the interlayer separation. We expect the marked
decrease in the spin polarization with bias to translate into an increase of
the NMR\ relaxation rate with increasing bias.

\begin{figure}[tbph]
\includegraphics[scale=1]{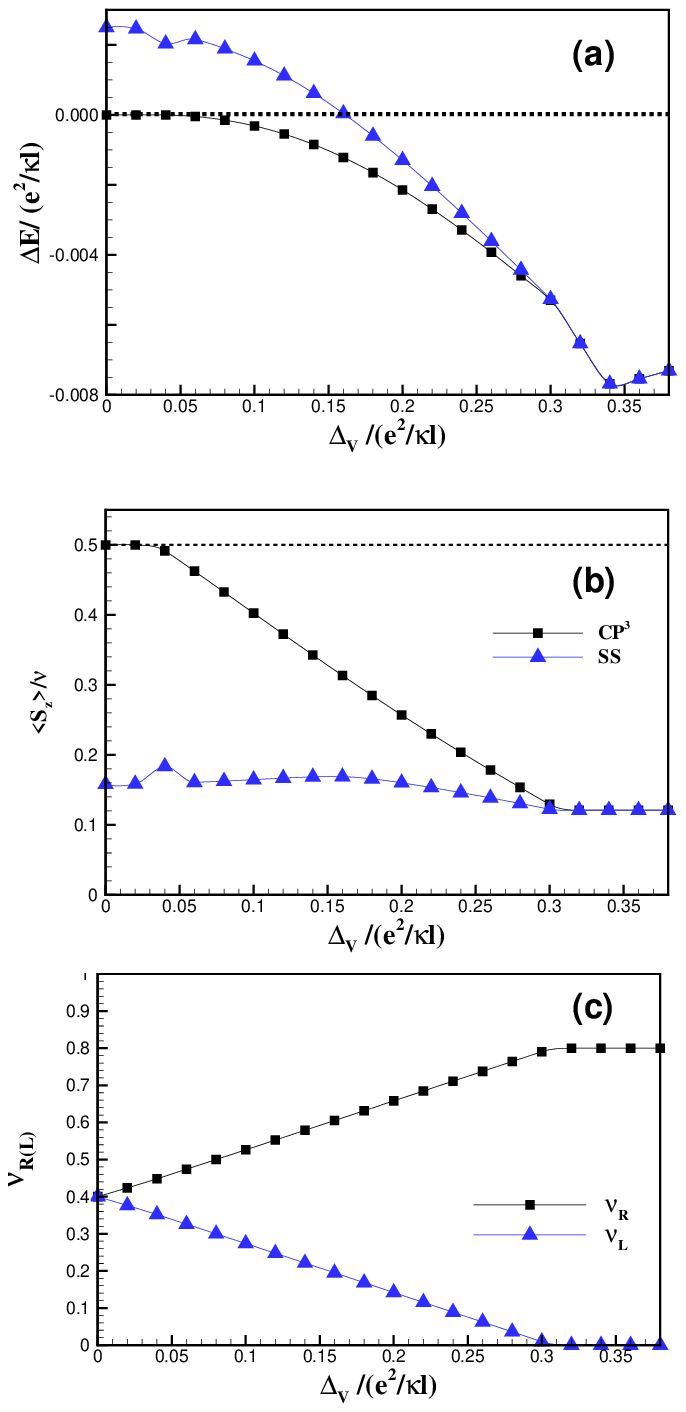}
\caption{(Color online). (a) Energy difference between the CP$^{3}$ or SS
states and the SPB state; (b) spin polarization per electron as a function
of bias for $\protect\nu =0.8,\Delta _{SAS}/\left( e^{2}/\protect\kappa \ell
\right) =0.0002,$ and $\Delta _{Z}/\left( e^{2}/\protect\kappa \ell \right)
=0.01$; and (c) filling factor in the right and left wells in the CP$^{3}$
state.}
\label{figure9}
\end{figure}

\section{Discussion and Conclusion}

Our numerical calculations show that crystals involving spin and/or
pseudospin textures are likely candidates for the ground state of the 2DEG
in a bilayer quantum Hall system around filling factor $\nu =1.$ At small
tunneling and for $\nu <1$, we find intertwined spin and pseudospin textures
(CP$^{3}$ crystal) with a spin polarization that is strongly interlayer
dependent while at higher tunneling, a symmetric skyrmion state with fully
polarized pseudospins or another type of spin-pseudospin state minimizes the
energy.

As mentioned in our Introduction, a Skyrmion crystal has an extra gapless
spin mode in the crystal phase (and possibly in some overdamped form in a
Skyrme liquid state) that is believed to be responsible for the rapid
nuclear spin relaxation observed in the experiments\textit{.}\cite%
{cotegirvinprl}. This extra Goldstone mode is present both in the SS and CP$%
^{3}$ crystal states that we studied in this paper but not, in the SPB state.%
\cite{cotenmr} To make a direct comparison with the experiments of~Spielman 
\textit{et al}. and Kumada \textit{et al., }it is necessary to compute the
NMR\ relaxation rate. Results of such calculations will be presented
elsewhere. \cite{cotenmr} We can expect, however, that the relaxation rate
will be proportional to the in-plane spin polarization so that the behavior
of spin polarization $S_{z}$ should be an indication of the behavior of the
relaxation time $T_{1}$. The rapid change in the spin polarization that we
found in the CP$^{3}$ crystal state (very small $\Delta _{SAS}$) for filling
factor around $\nu =1$ may explain the rapid change in the NMR\ relaxation
rate measured in the experiment of Spielman \textit{et al}. which was
carried on at almost zero tunneling and for a Zeeman coupling which is
approximately that indicated in Fig. 4.

\begin{figure}[tbph]
\includegraphics[scale=1]{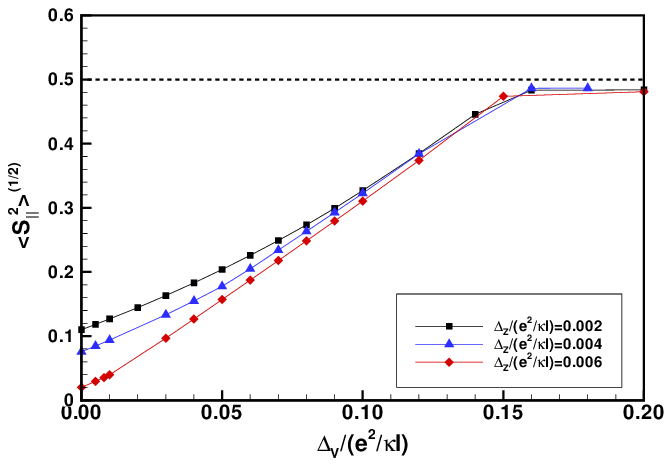}
\caption{(Color online). Average in-plane spin polarization in the CP$^{3}$
crystal state as a function of applied bias for $\protect\nu =0.8,\Delta
_{SAS}/\left( e^{2}/\protect\kappa \ell \right) =0.0002$ and $d/\ell =1.0$}
\label{figure10p}
\end{figure}

Our Hartree-Fock calculation indicates that the ground state at higher
tunneling is a SS state instead of a CP$^{3}$ crystal. In this case, the
spin polarization $S_{z}$ varies much less rapidly with filling factor than
for CP$^{3}$ crystal. Moreover, the spin polarization does not depend much
on the interlayer separation as can be seen, for example, in Figs. 3. The
results of Kumada \textit{et al}. showing a rapid change in the NMR\
relaxation rate as well as a strong dependence on the interlayer separation
would be more readily explained by a CP$^{3}$ crystal state than by the SS
state that we find. This is also true for their measurement of the
relaxation rate in the presence of an applied bias.

Fig. 10 illustrates the in-plane component for various Zeeman couplings as a
function of bias, which we believe is a measure of the NMR relaxation rate,
for small $\Delta _{SAS}$. The evident continuous behavior is reminiscent of
the Kumada results. We speculate that the effects of finite well width,
Landau level mixing, and possibly disorder, all not included in our
calculations, may stabilize the CP$^{3}$ state at higher tunneling than was
found for our idealized system.

In conclusion, we have studied textured quantum Hall states in bilayer
systems including the spin degree of freedom, and have shown under
appropriate circumstances that mixed spin-pseudospin textures appear as the
groundstate.

\section{Acknowledgements}

This work was supported by a research grant (for R.C.) and graduate research
grants (for J.B.) both from the Natural Sciences and Engineering Research
Council of Canada (NSERC). H.A.F. acknowledges the support of NSF through
Grant No. DMR-0454699. K. J. M. and B. R. acknowledge the support of NSF
Grants No. DMR-0080054, EPS-9720651 and DMR-0454699. Also B. Roostaei thanks
KITP Santa Barbara as part of this work was performed there. R. C. and J. B.
thank the R\'{e}seau qu\'{e}b\'{e}cois de calcul haute performance for
computer time.


\begin{thebibliography}{99}
\bibitem{dassarmalivre} S. Das Sarma and A. Pinczuk, Perspectives in Quantum
Hall Effects: Novel Quantum Liquids in Low-Dimensional Semiconductor
Structures (Wiley, 1996).

\bibitem{ezawalivre} Z.F. Ezawa, \textit{Quantum Hall Effects:\ Field
Theoretical Approach and Related Topics }(World Scientific, 2000). In this
book, the magnetic field $\mathbf{B}=-B_{0}\widehat{\mathbf{z}}$ while our
convention in this paper is that $\mathbf{B}=B_{0}\widehat{\mathbf{z}}$.
Thus, our skyrmion has opposite vorticity and real charge from the solution
given in this book.

\bibitem{sawadacp3} A. Sawada, D. Terasawa, N. Kumada, M. Morino, K.
Tagashira, Z.F. Ezawa, K. Muraki, T. Saku, and Y. Hirayama, Physica E 
\textbf{18}, 118 (2003); D. Terasawa, M. Morino, K. Nakada, S. Kozumi, A.
Sawada, Z.F. Ezawa, N. Kumada, K. Muraki, T. Saku and Y. Hirayama, Physica E 
\textbf{22}, 52 (2004); A. Sawada, Z.F. Ezawa, H. Ohno, Y. Horikoshi, A.
Urayama, Y. Ohno, S. Kishimoto, F. Matsukura, and N. Kumada, Phys. Rev. B 
\textbf{59}, 14888 (1999).

\bibitem{spielmanprl} I. B. Spielman, L. A. Tracy, J. P. Eisenstein, L. N.
Pfeiffer, and K.W. West, Phys. Rev. Lett. \textbf{94}, 076803 (2005).

\bibitem{kumadaprl} N. Kumada, K. Muraki, K. Hashimoto, and Y. Hirayama,
Phys. Rev. Lett. \textbf{94}, 096802 (2005); N. Kumada, K. Muraki, and Y.
Hirayama, Physica E (submitted), 2005.

\bibitem{sondhi} S.L. Sondhi, A. Karlhede, S.A. Kivelson, and E.H. Rezayi,
Phys. Rev. B \textbf{47}, 16419 (1993).

\bibitem{barret} S. E. Barrett, G. Dabbagh, L. N. Pfeiffer, K. W. West, and
R. Tycko, Phys. Rev. Lett. \textbf{74}, 5112 (1995).

\bibitem{schmeller} A. Schmeller, J. P. Eisenstein, L. N. Pfeiffer, and K.
W. West, Phys. Rev. Lett. \textbf{75}, 4290 (1995).

\bibitem{murphy} S. Q. Murphy, J. P. Eisenstein, G. S. Boebinger, L. N.
Pfeiffer, and K. W. West, Phys. Rev. Lett. \textbf{72}, 728 (1994).

\bibitem{kunyangci} Kun Yang, K. Moon, L. Zheng, A. H. MacDonald, S. M.
Girvin, D. Yoshioka, and Shou-Cheng Zhang, Phys. Rev. Lett. \textbf{72}, 732
(1994).

\bibitem{breycristalbimerons} L. Brey, H.A. Fertig, R. C\^{o}t\'{e}, and
A.H. MacDonald, Phys. Rev. B \textbf{54}, 16888 (1996).

\bibitem{rajaramancp3} S. Ghosh and R. Rajaraman, Phys. Rev. B \textbf{63},
035304 (2001).

\bibitem{ezawasu4} Z.F. Ezawa, Phys. Rev. Lett. \textbf{82}, 3512 (1999);
Z.F. Ezawa and G. Tsitsishvili, Phys. Rev. B \textbf{70}, 125304 (2004);
Z.F. Ezawa, Physica B \textbf{463}, 294-295 (2001); Z.F. Ezawa and K.
Hasebe, Phys. Rev. B \textbf{65}, 075311 (2002).

\bibitem{tyckonmr} R. Tycko, S. E. Barret, G. Dabbagh, L. N. Pfeiffer, and
K. W. West, Science \textbf{268}, 1460 (1995).

\bibitem{cotegirvinprl} R. C\^{o}t\'{e}, A.H. MacDonald, L. Brey, H.A. Fertig, S.M. Girvin, and H. T. C.  Stoof, Phys. Rev. Lett, \textbf{78}, 4825 (1997).

\bibitem{breyprl} L. Brey, H.A. Fertig, R. C\^{o}t\'{e} and A.H. MacDonald,
Phys.Rev. Lett. \textbf{75}, 2562 (1995).

\bibitem{cotemethode} R. C\^{o}t\'{e} and A. H. MacDonald, Phys. Rev. B 
\textbf{44}, 8759 (1991).

\bibitem{fertigdispersion} H. A. Fertig, Phys. Rev. B \textbf{40}, 1087
(1989).

\bibitem{biastheory} Y. N. Joglekar and A. H. MacDonald, Phys. Rev. B 
\textbf{65}, 235319 (2002).

\bibitem{biasexperiment} I. B. Spielman, M. Kellogg, J. P. Eisenstein, L. N.
Pfeiffer, and K. W. West, Phys. Rev. B \textbf{70}, 081303(R) (2004).

\bibitem{macdobible} K. Moon, H. Mori, K. Yang, S.M. Girvin, A.H. MacDonald,
L. Zheng, D. Yoshioka and S. C.  Zhang, Phys. Rev. B \textbf{51}, 5138 (1995);
K. Yang, K. Moon, L. Belkhir, H. Mori, S.M. Girvin, A.H. MacDonald, L.
Zheng, and D. Yoshioka, Phys. Rev. B \textbf{54}, 11644 (1996).

\bibitem{fertigskyrmion} H. A. Fertig, L. Brey, R. C\^{o}t\'{e}, and A.H.
MacDonald, Phys. Rev. B50, 11018 (1994).

\bibitem{cotegroup1} R. C\^{o}t\'{e}, M. Boissonneault and M. Dion.
Unpublished.

\bibitem{breymeron} L. Brey, H.A. Fertig, R. C\^{o}t\'{e}, and A.H.
MacDonald, Physica Scripta, T \textbf{66}, 154 (1996).

\bibitem{nazarov} Y. V. Nazarov and A. V. Khaetskii, Phys. Rev. Lett. 
\textbf{80}, 576 (1998).

\bibitem{ezawabimeron} Z.F. Ezawa, Phys. Rev. B \textbf{55}, 7771 (1995).

\bibitem{rajaramanbimeron} S. Ghosh and R. Rajaraman, International Journal
of Modern Physics B \textbf{12}, 2495 (1998); ibid. p. 37.




\bibitem{cotenmr} R. C\^{o}t\'{e} et al. Unpublished.
\end{thebibliography}
\end{document}